\documentclass[10pt,twocolumn,aps,pra,superscriptaddress]{revtex4-2}
\usepackage{graphicx,subfigure,physics,dsfont,amsmath,appendix,amssymb,mathrsfs}
\usepackage{blkarray}
\usepackage[utf8]{inputenc}
\usepackage[normalem]{ulem}               
\usepackage{orcidlink}
\usepackage{duckuments}
\usepackage[dvipsnames]{xcolor}           
\usepackage{times,txfonts}
\usepackage{hyperref}
\usepackage{color}
\usepackage{todonotes}
\definecolor{mycolor}{rgb}{0.188, 0.196, 0.654}           
\definecolor{indiagreen}{rgb}{0.01, 0.43, 0.01}
\definecolor{indigo}{rgb}{0.29, 0.0, 0.51}
\hypersetup{
    colorlinks=true,
    linkcolor={red!50!black},
    citecolor={blue!80!black},
    urlcolor={purple!80!black},
    linktoc=all
}

\begin{document}

\title{Dynamics of two bosons with density-assisted tunnelling in a one-dimensional tilted lattice }

\author{Muhammad S. Hasan\,\orcidlink{0000-0002-2549-8656}}
\thanks{Equal contribution}
\email[]{mshasan.official@gmail.com}
\affiliation{Institute of Atomic and Molecular Sciences, Academia Sinica, Taipei 10617, Taiwan}

\author{Vikash Mittal\,\orcidlink{0000-0002-4384-6992}}
\thanks{Equal contribution}
\email[]{vikashmittal.iiser@gmail.com}
\affiliation{Institute of Physics, Polish Academy of Sciences, Aleja Lotnikow 32/46, PL-02668 Warsaw, Poland}

\author{Tomasz Sowi\'nski\,\orcidlink{0000-0002-7970-4371}}
% \email[]{tomasz.sowinski@ifpan.edu.pl}
\affiliation{Institute of Physics, Polish Academy of Sciences, Aleja Lotnikow 32/46, PL-02668 Warsaw, Poland}

\begin{abstract}
We investigate the dynamics of two bosons in a tilted one-dimensional optical lattice with both standard and density-assisted tunnelling. The relative sign of the two tunnelling processes strongly affects the Bloch dynamics, leading to distinct reinforcing and competing regimes. We examine how the tunnelling processes generate correlations and find that, in the absence of onsite interactions, the early-time entanglement growth follows the same scaling with the density-assisted tunnelling amplitude. Finite onsite interaction changes this picture by breaking the symmetry between the reinforcing and competing regimes and enhancing the correlation differently in two regimes. This shows how the interplay between the two tunnelling mechanisms can change dyanmics and entanglement in a controlled few-body setting. Particularly, at equal tunnelling amplitudes, the doublon and singlon sectors decouple completely for arbitrary onsite interaction, lattice tilting, and initial state.
\end{abstract}

\maketitle
%------------------------------
\section{Introduction}
%------------------------------
Ultracold atoms in optical lattices have been a cornerstone of quantum simulations owing to high controllability and tunability eventually providing a platform for exploring many-body physics \cite{Greiner:2002, Lewenstein:2007, Bloch:2008, Jordens:2008, Gross:2017, Schafer:2020, Takahashi:2022}. In the simplest scenario, these systems are described by the celebrated Bose--Hubbard model, which treats two-body interactions by retaining only the onsite interaction while neglecting all other terms \cite{Fisher:1989, Jaksch:1998, Kuhner:1998, Jaksch:2005, Isart:2007}. While this approximation is proven useful in many contexts, it becomes restrictive when considering that finite-range interactions extending over multiple sites can give rise to interesting tunnelling phenomena that has an effect on the overall dynamics of the system. Even for the simplest form of contact interactions, when lattice is not deep enough, a tight-binding model can lead to additional correlated tunnelling terms in the effective lattice Hamiltonian \cite{Dutta:2011, Bissbort:2012, Luhmann:2012, Lacki:2013}. Such correlated tunnelling processes, together with contributions from higher bands, have been shown to account for shifts in the superfluid-Mott insulator transition in Bose--Fermi and Bose--Bose mixtures and to modify the band structure in optical lattices \cite{Gunter:2006, Ospelkaus:2006, Heinze:2011, Maik:2013, Lacki:2013}. One such effective process is the density-assisted tunnelling, which describes the particle tunnelling to a neighbouring site induced by the presence of particles at that site \cite{Jurgensen:2012, Maik:2013, Jurgensen:2014, Wang:2019}.

The motion of particles in an optical lattice can naturally be viewed as a  quantum walk, where coherent tunnelling between neighbouring sites leads to the spreading of a particle over multiple lattice sites \cite{Dur:2002, Kempe:2003, Eckert:2005, Chandrashekar:2008, Karski:2009, Witthaut:2010, Andraca:2012,  Wang:2014, Preiss:2015}. From this perspective, it provides a direct way to study transport properties. Quantum walks with interacting bosons have revealed the emergence of correlations between particles and the formation of strongly correlated transport regimes \cite{Wang:2014, Preiss:2015}. These developments make the two-particle quantum walk a natural setting for studying how different tunnelling processes and mutual interactions modify the motion and correlations of interacting particles. 

Given its significance, the role of density-assisted tunnelling in the dynamics of interacting particles deserves further exploration. Several previous works have investigated interacting particles using an extended Bose--Hubbard model, some even considering multi-band tunnelling, yet the majority of such studies have either neglected the process of density-assisted tunnelling altogether or focussed exclusively on the ground-state phase diagrams \cite{Mazzarella:2006, Sowinski:2012, Jurgensen:2012, Luhmann:2012, Dariusz:2017, Sarkar:2020, Chanda:2025}. In this work, we fill this gap by analysing the dynamics of two interacting bosons in an optical lattice with explicit inclusion of density-assisted tunnelling. This process has been observed experimentally in tilted lattices \cite{Jurgensen:2014, Meinert:2016}. We focus on pushing the understanding of the interplay of the two tunnelling processes on the system and therefore we choose a system with only two bosons. As explained in \cite{Sowinski:2012}, the relative strength of two tunnelling processes can be varied over a wide range. Therefore, in our theoretical analysis, we treat them as independently controllable parameters.

The manuscript is organised as follows. Section~\ref{sec:model} introduces the extended Bose--Hubbard Hamiltonian, and numerical methods used throughout the work. We first consider the non-interacting system in Secs.~\ref{sec:U0_dynamics}--\ref{sec:earlytime_entanglement} and investigate the effect of the two tunnelling processes focussing particularly on the correlations generated by density-assisted tunnelling. In Sec.~\ref{sec:onsite_interactions}, we introduce a finite onsite interaction and study how it changes the correlations and in Sec.~\ref{sec:lambda_0.5} we discuss a special case of the destructive interference of the two tunnelling mechanisms. We conclude in Sec.~\ref{sec:conclusion} by summarising the our results and discussing possible extensions. 

%------------------------------
\section{Model and Methods}
\label{sec:model}
%------------------------------
We consider two spinless bosons moving in a one-dimensional optical lattice occupying the lowest Bloch band. We assume that the system considered can be described with the Hamiltonian of the form \cite{Jaksch:1998} 
\begin{align} 
\hat{H} &= -J \sum_i \hat{a}^\dagger_i(\hat{a}_{i-1} + \hat{a}_{i+1})
+ F \sum_i (i-i_0)~\hat{a}^\dagger_i\hat{a}_i  \nonumber \\
&+ \frac{U}{2} \sum_i \hat{n}_i(\hat{n}_{i}-1) +T\sum_i\bigg(\hat{a}^\dagger_i \left(\hat{n}_i+\hat{n}_{i-1}\right)\hat{a}_{i-1} + \mathrm{H.c.}\bigg)\;,
\label{eq:finalHamiltonian}
\end{align}
where operator $\hat a_i^{\dagger}(\hat a_i)$ creates (annihilates) a boson at site $i$ and obeys the bosonic commutation relation, while $\hat n_i = \hat a_i^{\dagger}\hat a_i$ is the number operator at site $i$. The first term describes the nearest neighbour hopping of bosons with amplitude $J$, while the second term describes the linear tilt of strength $F$ that breaks the translational symmetry of the lattice. The centre of the lattice is taken as $i_0$ which is arbitrary and does not affect the dynamics. The third term accounts for the onsite interaction with strength $U$. The last term captures the density-assisted tunnelling with amplitude $T$ where the hopping depends on the presence or absence of the particle in neighbouring sites. To facilitate the analysis, we rescale all interaction strengths with respect to the tilt strength $F$ such that the energy and time are measured in units of $F$ and $\hbar/F$ respectively. Additionally, we parameterise the tunnellings $J$ and $T$ as
\begin{equation}
    \begin{aligned}
        J(\lambda) = J_0 \cos \left( \lambda \dfrac{\pi}{2}\right), \;\;\;\; T(\lambda) = T_0 \sin \left( \lambda \dfrac{\pi}{2}\right)
    \end{aligned}
\end{equation}
which allows us to perform analysis with a single parameter $\lambda$. For the reference we set $J_0 = T_0 = 5F$. For two bosons, the Fock basis can be split into two subspaces, the doublon  subspace, spanned by the states $|D_k\rangle = \frac{(\hat{a}^\dagger_k)^2}{\sqrt{2}}|0\rangle$ and the singlon subspace, spanned by the states $|S_{k,l} \rangle = \hat{a}^\dagger_k\hat{a}^\dagger_l|0\rangle, k<l$. This decomposition will be useful in analysing the dynamics throughout the paper.

We assume that the initial state for the time evolution is a Gaussian wavepacket of two bosons of the form \cite{Dariusz:2017} 
\begin{equation}
    \ket{\Psi_0} = \mathcal{N} \bigg( \sum_i e^{-{(i-i_0)^2/2\sigma^2}} \hat a_i^\dagger \bigg)^2 \ket{0}\;,
    \label{eq:gauss_initial}
\end{equation}
where $\sigma$ is the width of the wavepacket and $\mathcal{N}$ is the normalisation constant. The time evolved state $|\Psi(t)\rangle$ is obtained by expanding the initial state in the eigenbasis of the Hamiltonian $\hat H$
\begin{equation}
    |\Psi(t)\big\rangle = \sum_k \big\langle\psi_k\big|\Psi_0\big\rangle e^{-iE_k t} |\psi_k\big\rangle\;,
    \label{eq:time_ev}
\end{equation}
where $E_k$ and $|\psi_k\rangle$ are the eigenenergies and eigenstates of the many-body Hamiltonian respectively. To obtain them, we restrict the size of the lattice to sufficiently large number of lattice sites $M$ (in our case $M=101$) and use the exact diagonalisation to numerically diagonalise the the many-body Hamiltonian in the Fock space basis, which is constructed from the occupation number of the bosons at each lattice site \cite{Zhang:2010, Raventos:2017}.

We start the analysis by focussing on the density distribution of bosons on the lattice, which is straightforward to calculate using the time-evolved state $|\Psi(t)\rangle$ as
\begin{equation}
    n_i^{tot}(t) = \big\langle\Psi(t)|\hat{a}_i^\dagger \hat{a}_i|\Psi(t)\big\rangle\;.
\end{equation}
This total density can be thought of as contribution from singlons and doublons. The doublon density, which is the probability of finding both the bosons at the same site, is given by
$n^{(2)}_i(t) = 2~\big|\big\langle D_i|\Psi(t)\big\rangle\big|^2$ while the singlon density can be calculated as the difference $n^{(1)}_i(t) = n_i^{tot}(t) - n^{(2)}_i(t)$. This decomposition allows us to distinguish between different spatial configurations of the bosons and provides insight into the dynamics of different contributions as discussed in the following sections. 
%%%%%%%%%%%%%%%%%%%%%%%%%%%%%%%%%%%%%%%%%%%%%%%%%%%%%%
%---------------Results-------------
%%%%%%%%%%%%%%%%%%%%%%%%%%%%%%%%%%%%%%%%%%%%%%%%%%%%%%%
\section{Dynamics without on-site interactions}
\label{sec:U0_dynamics}
\begin{figure}[h!]
    \centering
    \includegraphics[width=\linewidth]{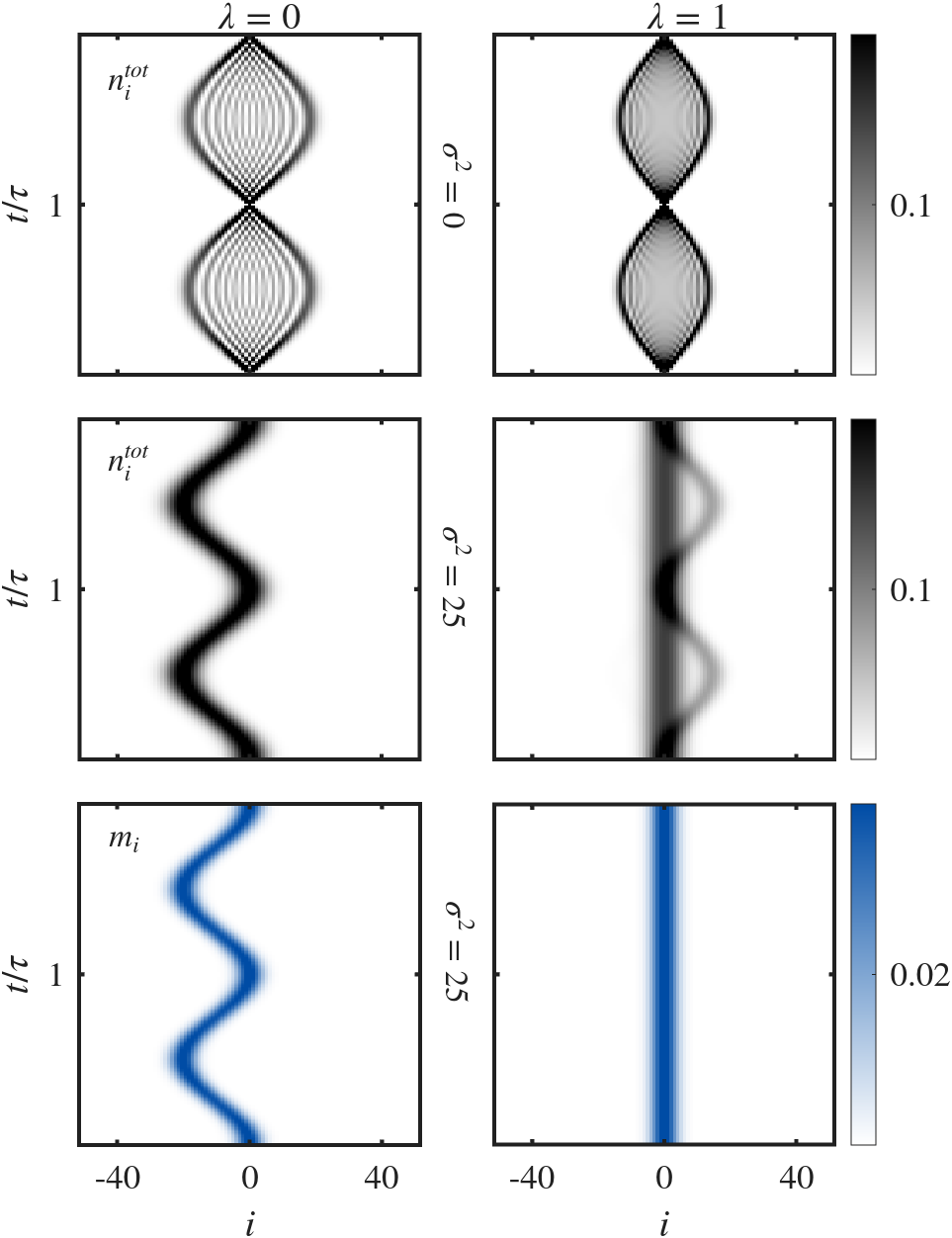}
     \caption{Time-evolved density distributions $n_i^{tot}(t)$ of two non-interacting bosons ($U=0$) for standard hopping only $\lambda = 0$, (left column) and density-assisted tunnelling only $\lambda=1$ (right column). The top row shows both bosons initially localised at the centre of the lattice with $\sigma^2=0$. The middle row shows Gaussian wave packet with width $\sigma^2=25$. In the bottom row, the time evolved density $m_i(t)$ of the superposition of states where the bosons separated by one empty site is shown. The time evolution is performed over $t=[0, 2\tau]$.}
    \label{fig:U0_density_adj}
\end{figure}

%%%%%%%%%%%%%%%%%%%%%%%%%%%%%%%%%%%%%%%%%%%%%%%%%%%%%%%%
%----------------U=0 section-----------
To understand the dynamics of the two-boson system governed by the Hamiltonian \eqref{eq:finalHamiltonian} and to benchmark the standard tunnelling, we first consider the case when the onsite interaction is zero ($U=0$) and the density-assisted tunnelling is switched off ($\lambda=0$). The left column of Fig.~\ref{fig:U0_density_adj} shows the time evolution of the total density profile of the two bosons in the presence of the uniform tilt $F$. The external potential results in the well-known phenomenon of the Bloch oscillations that has been studied extensively across various systems both theoretically and experimentally \cite{Bloch:1929, Dahan:1996, Kolovsky:2004, Morsch:2006, Dias:2007, Khomeriki:2010, Morsch:2016, Hasan:2022, Rabec:2025}. 
\begin{figure*}
    \centering
        \includegraphics[width=\linewidth]{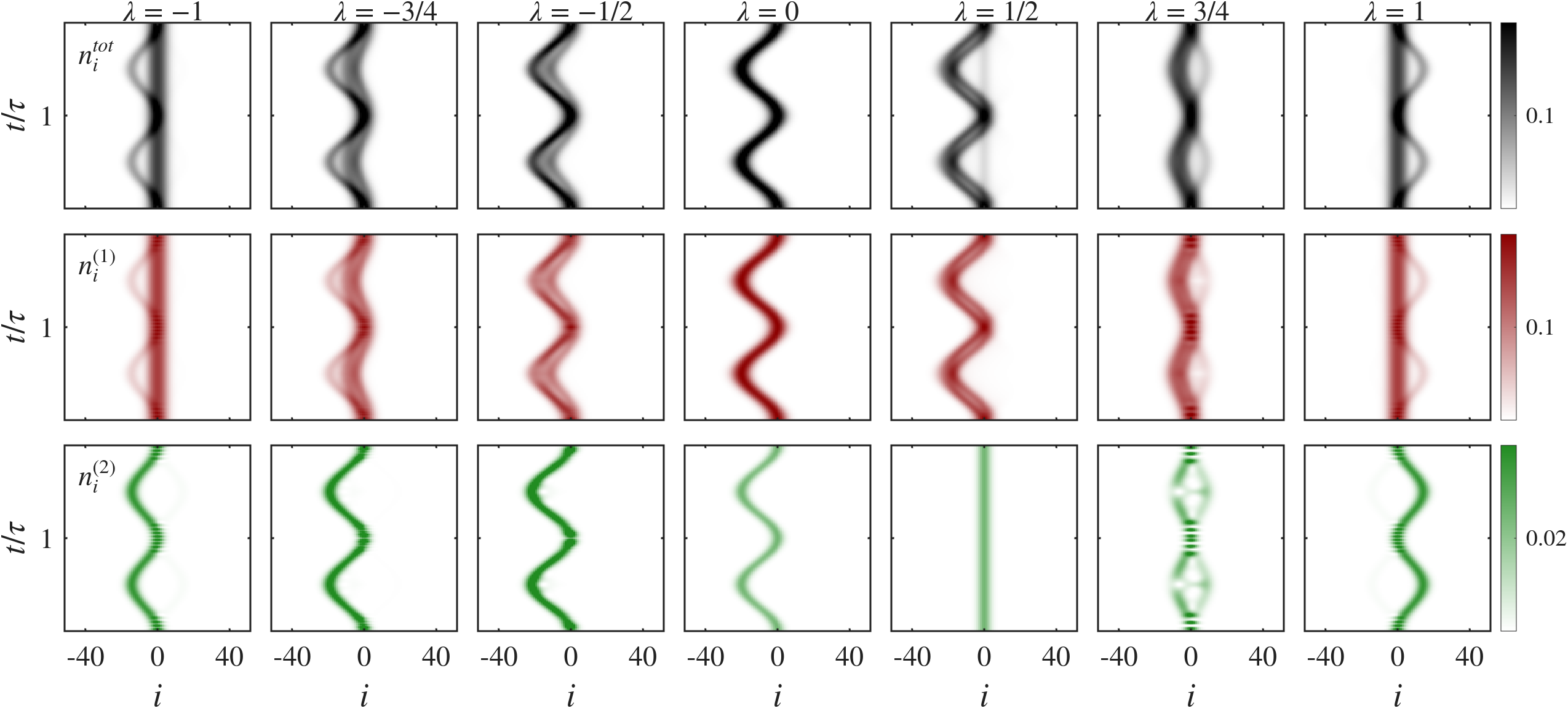}
    \caption{Time-evolved density distributions for non-interacting bosons, ($U=0$) prepared initially in a Gaussian state \eqref{eq:gauss_initial} with $\sigma^2=25$. Top row shows total density $n_i^{tot}$. Singlon density $n_i^{(1)}$ and doublon density $n_i^{(2)}$ are shown in middle and bottom rows, respectively. At $\lambda=0$, all three densities exhibit regular Bloch oscillations with period $\tau  = 2\pi/F$. At $\lambda=+1/2$, the doublon density is completely frozen due to destructive interference between $J$ and $T$, while singlons remain oscillatory. The reinforcing regime $\lambda < 0$ produces enhanced oscillation amplitudes, whereas the competitive regime $\lambda > 0$ leads to reduced amplitudes and the emergence of a non-oscillating background. At $\lambda = \pm 1$, the dynamics is governed entirely by the density-assisted tunnelling. All simulations are shown for two Bloch periods ($t = [0,2\tau]$).}
    \label{fig:U0_density_all}
\end{figure*}
In the top row of Fig.~\ref{fig:U0_density_adj} we show the case when the two bosons initially occupy the same site ($\sigma^2 \rightarrow 0$). In the middle and the bottom rows, the initial state is a Gaussian of width $\sigma^2 =25$ extending over multiple lattice sites and hence we see different dynamics. The Bloch oscillations remain intact and we can see here that the oscillation frequency is the same for any $\sigma$. This is because the time period of the oscillation $\tau = \frac{2 \pi \hbar}{F}$ only depends on the external linear potential strength \cite{Dahan:1996, Kolovsky:2004, Morsch:2006}. We recover the results of the dynamics of a single particle in a lattice subjected to tilt \cite{Kolovsky:2003, Jaczewski:2024}.

When the standard hopping $J$ is switched off and only the density-assisted tunnelling remains active ($\lambda=1$), the oscillation amplitude is reduced, though the Bloch oscillations remain, as shown in the right column of Fig.~\ref{fig:U0_density_adj}. For bosons initially on the same site (top row), the concentric density structures seen for $\lambda=0$ are suppressed. For the Gaussian initial state with $\sigma^2=25$, (middle row), the oscillations are accompanied by a static background, visible as the straight line in the density profile.

This phenomenon can be understood by considering the projection operator $\hat{P}$ defined as
\begin{equation}
    \hat {P} = \sum_i \hat{a}_i^\dagger \hat{a}_{i+2}^\dagger \hat{a}_{i+2} \hat{a}_i = \sum_i |S_{i,i+2}\rangle \langle S_{i,i+2}| \;.
\end{equation}
This operator projects onto combinations of the Fock states where the bosons are occupying sites $i$ and $i+2$. It can be shown straightforwardly, that in the case of two bosons, it projects to the kernel of any density tunneling process, \textit{i.e.},
\begin{equation}
\hat{a}_i^\dagger(\hat{n}_{i} + \hat{n}_{i-1})\hat{a}_{i-1}\hat {P} \equiv 0\;.
\end{equation}
Therefore, the density of such states
\begin{equation}
    m_i(t) = \langle\Psi(t)|\hat{P}^\dagger \hat{a}_i^\dagger \hat{a}_i \hat{P}|\Psi(t)\rangle\;,
\end{equation}
remains spatially frozen because the density-assisted tunnelling results in zero contribution within this subspace. This is clearly shown in the bottom row of the right column of Fig.~\ref{fig:U0_density_adj}. Even though the tilt drives the particles to exhibit Bloch oscillations for other states, the states where the particles are separated by one empty site exhibit no such oscillation when only the density-assisted tunnelling is present. For completeness, we show on the bottom-left panel of Fig.~\ref{fig:U0_density_adj} the case of standard tunnelling ($\lambda=0$), the density of such separated states follow the same pattern as of the total density.
%%%%%%%%%%%%%%%%%%%%%%%%%%%%%%%%%%%%%%%%%%%%%%%%%%%%%%
%{Singlon and Doublon contributions}
%%%%%%%%%%%%%%%%%%%%%%%%%%%%%%%%%%%%%%%%%%%%%%%%%%%%%%
\section{Interplay of tunnelings}
\label{sec:interplay_of_Jand_T}
The interplay of $J$ and $T$ on the dynamics depends on their relative signs. This cooperation and competition of the hopping mechanisms can be clearly seen in Fig.~\ref{fig:U0_density_all}, where we show the time-evolved spatial density distribution for the total density $n_i^{tot}$ (top row), singlon density $n_i^{(1)}$ (middle row) and doublon density $n_i^{(2)}$ (bottom row), for various values of $\lambda$. 

%Let us first consider the regime of vanishing density-assisted tunnelling ($\lambda=0$) where the system reduces to the standard tilted lattice with nearest-neighbour hopping. In this limit, the total, singlon, and doublon densities exhibit regular Bloch oscillations, characteristic of tunnelling in a periodic potential under a constant force. While the singlon density shows larger intensity, the doublon contribution is reduced, though, both oscillate in phase, reflecting the uniform and interaction-free nature of the dynamics.

An interesting regime emerges at $\lambda =\pm 1/2$ where the interplay between $J$ and $T$ leads to different dynamical behaviour depending on the sign of $\lambda$. For $\lambda=+1/2$, the total density develops a mixed structure consisting of a weakly oscillatory part superimposed on a non-oscillating background. In contrast, for $\lambda=-1/2$, the total density exhibits no such non-oscillating density. The singlon density remains oscillatory in both cases but displays subtle temporal patterns. For $\lambda=+1/2$, the singlon density periodically splits into two maxima with a dip in the middle, recombines at $t=\tau/4$ only to split again and recombine at $t=\tau/2$ and repeats in the next Bloch period as well. For $\lambda=-1/2$, a similar splitting--merging behaviour is observed but with a different pattern in density. A more drastic distinction is seen in the doublon behaviour. For $\lambda=+1/2$, the doublon density becomes completely static, indicating a dynamical freezing of doublons. This effect originates from the destructive interference between the two hopping mechanisms since $J$ and $T$ have opposite signs. For $\lambda=1/2$, this leads to exact cancellation of their contribution to doublon hopping. As a result the doublons experience no net tunnelling. In contrast, for $\lambda=-1/2$, this condition is lifted allowing doublons to tunnel and constructively interfere and exhibiting regular oscillations. This particular case has been shown to lead to non-ergodic dynamics in dipolar lattice bosons \cite{Aramthottil:2023}. We will revisit and discuss this case in detail in Sec.~\ref{sec:lambda_0.5}. 

For intermediate values of hoppings, such as $\lambda = -3/4$, the standard hopping and the density-assisted tunnelling reinforce each other, producing strong oscillations of the total density. In contrast, for $\lambda = +3/4$, the two mechanisms compete, resulting in reduced oscillation amplitude. In this case, the density also develops a component which hops on the opposite side of the lattice due to the partial cancellation of the two hopping mechanisms. For $\lambda=-3/4$, the singlon oscillation has larger amplitude while for $\lambda=+3/4$, the amplitude is noticeably smaller.

Finally, in the limit of vanishing standard hopping ($\lambda=\pm1$), the dynamics is fully governed by the density-assisted tunnelling. The total density can be viewed as the sum of two distinct contributions, an oscillatory density part and a static background. The latter arises from the contribution of the bosons separated by one empty site as discussed previously. The singlon density exhibits similar co-existing oscillating and static pattern with opposite direction of propagation due to the sign of $\lambda$.  Interestingly, the doublon density remains purely oscillatory and robust in both cases without the competition from the standard tunnelling $J$. 

\section{Quantitative analysis}
\label{sec:Hellinger}
\begin{figure}
    \centering
    \includegraphics[width=\linewidth]{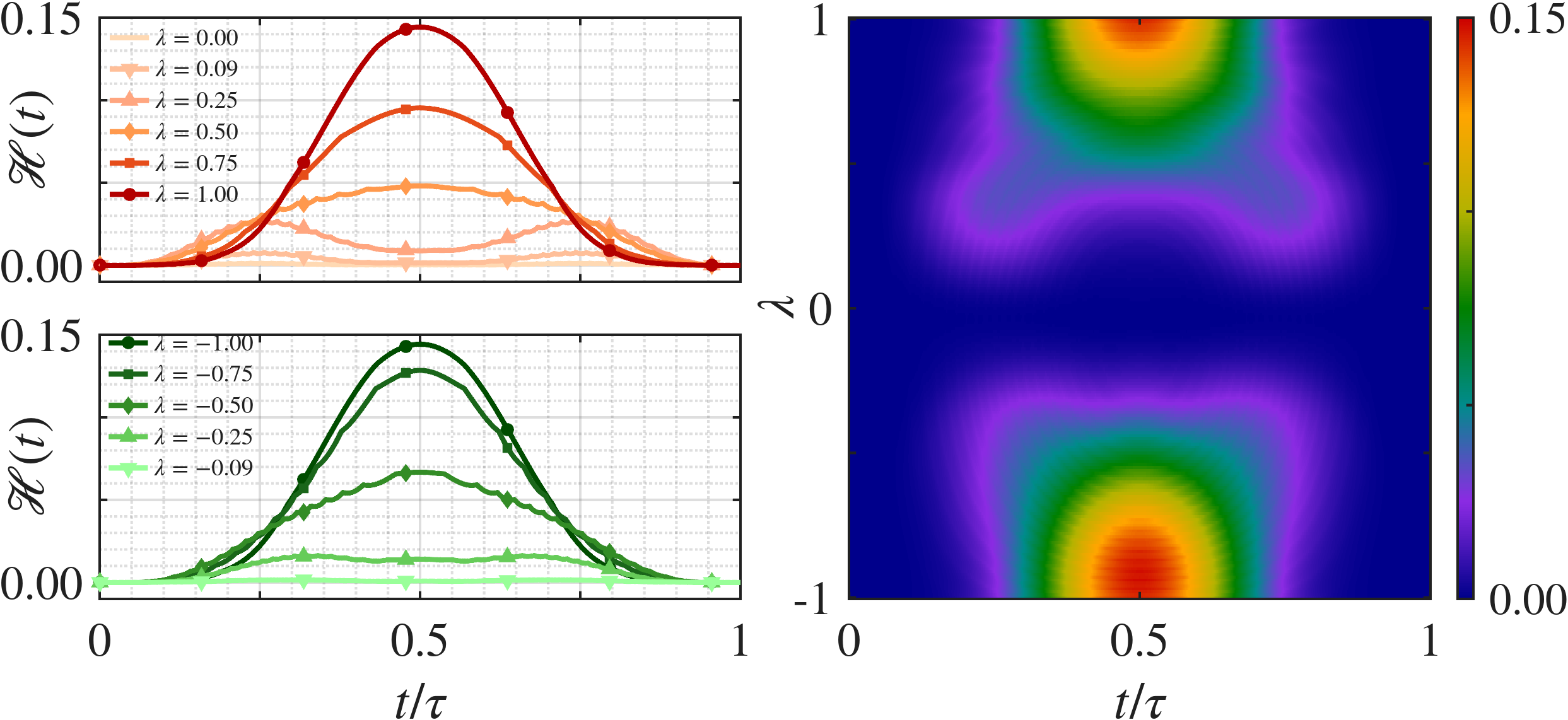}
        \caption{Hellinger distance $\mathscr{H}(t)$ as a function of time and different $\lambda$. For $\lambda = 0$, $\mathscr{H}(t) = 0$, confirming no distortion of the density profile under standard hopping alone. For $0 < \lambda < 1/2$, a double-peak structure emerges with maxima near $t = \tau/4$ and $t = 3\tau/4$, which collapses into a single peak near $t = \tau/2$ for $\lambda \geq 1/2$. No double-peak structure is observed for $\lambda < 0$.}
    \label{fig:Hellinger_vs_lambda}
\end{figure}

%--------------------Hellinger-------------------------------
To quantify these effects beyond visual inspection, we use Hellinger distance $\mathscr{H}(t)$, as a primary tool of measure for the deviations in densities during the dynamics \cite{Vegelius:1986, Chung:1989}. This probabilistic metric, closely related to the Euclidean distance, is bounded $0 \leq \mathscr{H}(t) \leq 1$, and captures the internal structural changes of the density profile during the dynamics. For our purposes, we define the Hellinger distance as 
\begin{equation}
    \mathscr{H}(t) = 1 - \frac{1}{2}\sum_i \sqrt{n^{ini}_{i-s(t)} \cdot n_i^{tot}(t)}\;,
    \label{Eq:Hellinger}
\end{equation}
where $n_i^{tot}(t)$ denotes the temporal normalised density at site $i$ and time $t$, while $n_{i-s(t)}^{ini}$ is the reference density obtained by shifting the initial distribution $n_i^{ini}$ according to the temporal expectation value of the position, $s(t) = \sum_i i\cdot n_i^{tot}(t)$. The Hellinger distance thus quantifies how strongly the time evolved density deviates from its initial form under the competing influences of the standard tunnelling and the density-assisted tunnelling. The Hellinger distance $\mathscr{H}(t)= 0$ corresponds to the situation where the the wavepacket maintains its initial shape at time $t$, while $\mathscr{H}(t)=1$ indicates zero overlap between the initial density and the time-evolved density at time $t$. In Fig.~\ref{fig:Hellinger_vs_lambda} we show how the Hellinger distance changes with time for various values of $\lambda$. In the top left panel of Fig.~\ref{fig:Hellinger_vs_lambda}, we show the case when $\lambda>0$. When the standard tunnelling is only present ($\lambda=0$), $ \mathscr{H}(t) = 0$ indicating no changes in the density profile. For intermediate values $0<\lambda<1/2$ we see a double peak structure. The Hellinger distance reaches maximum around $t=\tau/4$ then decreases then reaches maximum again around $t=3\tau/4$. For $\lambda\geq0.5$ the double peak structure disappears and $\mathscr{H}(t)$ reaches maximum around $t=\tau/2$ owing to the maximum deformation in the density profile around these times.

In the regime $\lambda<0$, no double peak is observed. Hellinger distance essentially remains zero in the vicinity of $\lambda= 0$. For larger negative values of $\lambda$, the distance increases, reaches a maximum near $t=\tau/2$, and subsequently decreases. While this overall behaviour is similar to when $\lambda>0$, the magnitude of the Hellinger distance is larger for negative $\lambda$. The asymmetry between $+\lambda$ and $-\lambda$ in the magnitude of the Hellinger distance is a direct consequence of the density dependence of the density-assisted tunnelling. Because the density-assisted tunnelling depends on the local occupation number, the reinforcing regime ($\lambda<0$) generates a larger effective hopping for doubly occupied sites and therefore drives stronger redistribution of the density. For completeness, in right panel of Fig.~\ref{fig:Hellinger_vs_lambda} we show the Hellinger distance against time over one Bloch period for various values of $\lambda$. It can be noted that positive and negative $\lambda$ values result in a different pattern of dynamics and hence the effect is seen in the Hellinger distance.

%%----------------------entropy from density-assisted tunneling----
\section{Correlations induced by tunneling}
\label{sec:entropy_nonint}
Let us now examine how the correlations induced by the density-assisted are built in the system. Since the onsite interaction is turned off  ($U=0$), any correlations produced in the system are purely due to this tunnelling process  \cite{Kraus:2022}. To quantify them we calculate the von Neumann entropy \cite{Luigi:2008, Horodecki:2009, Dellanna:2022}
\begin{equation}
    \mathcal{S}(t) = -\mathrm{Tr}\left[\rho^{(1)}(t)\,\ln\rho^{(1)}(t)\right],
\end{equation}
where $\rho^{(1)}(t)$ is the reduced single-particle density matrix with elements $\rho^{(1)}_{ij}(t) = \frac{1}{2}\langle\Psi(t)| \hat a_i^\dagger \hat a_j|\Psi(t) \rangle$. The diagonal part of normalised $\rho^{(1)}_{ii}(t)$ corresponds directly to the density distribution $n_i^{tot}(t)$ while the off-diagonal elements $\rho^{(1)}_{ij}$ ($i\neq j$) measure the single-particle coherences between sites $i$ and $j$.
\begin{figure}
    \centering
    \includegraphics[width=\linewidth]{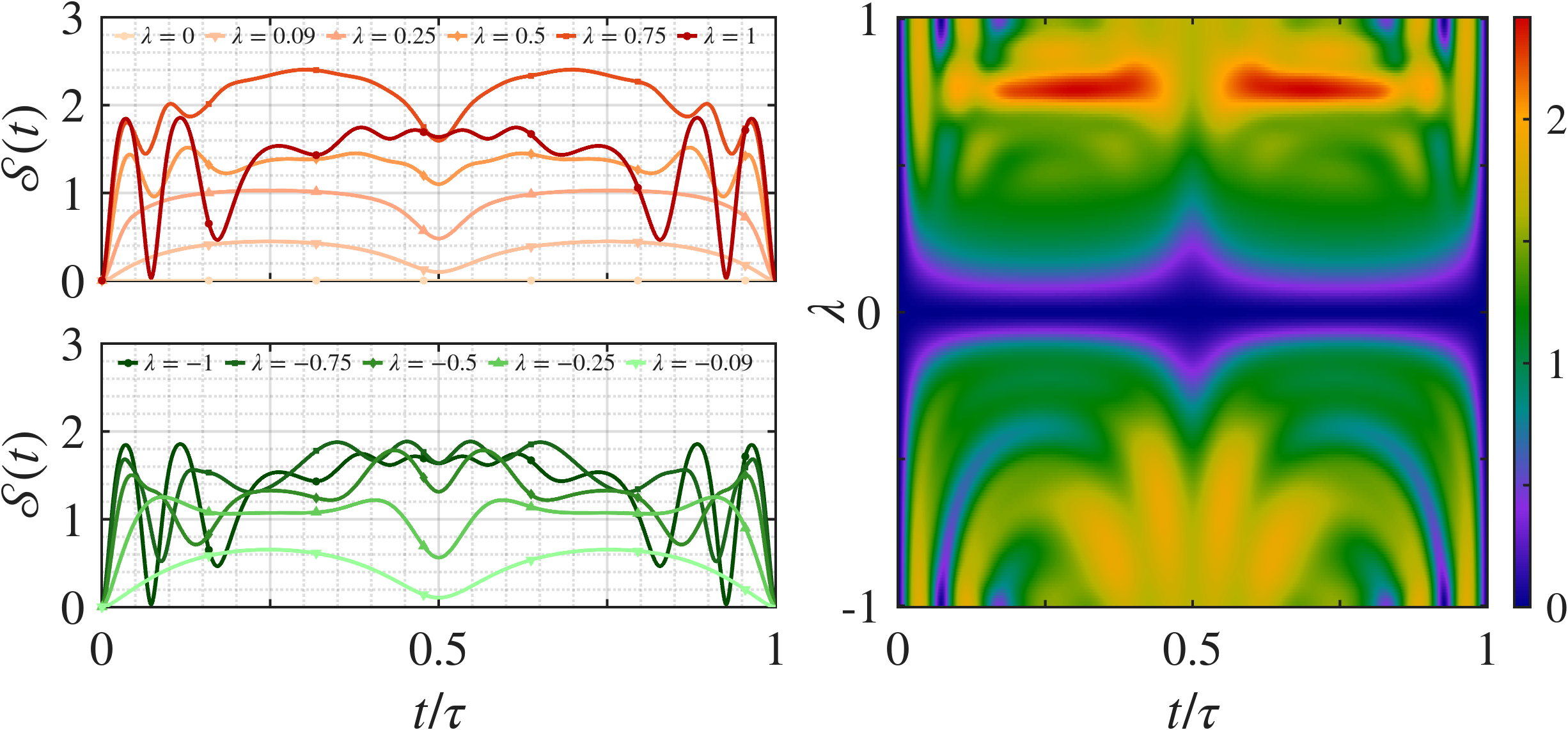}
    \caption{The von Neumann entropy $\mathcal{S}(t)$ as a function of time $t$ and parameter $\lambda$ for one Bloch period. At $\lambda = 0$, the entropy vanishes, confirming that standard hopping alone generates no inter-particle correlations. As $|\lambda|$ increases, the density-assisted tunnelling builds up correlations between particles, with the entropy being higher for negative $\lambda$. The entropy returns to zero at $t = \tau$ reflecting the periodicity of the Bloch dynamics.}
    \label{fig:entropy_zero_onsite}
\end{figure}

On right panel of Fig.~\ref{fig:entropy_zero_onsite} we show the entropy as a function of $\lambda$ and time for one Bloch period. For purely standard hopping ($\lambda = 0$), the Hamiltonian contains only single particle operators and the two bosons evolve independently generating no inter-particle correlation, \textit{i.e.} the entropy $\mathcal{S}(t) = 0$. As $|\lambda|$ increases, the density-assisted tunnelling comes into the picture and generates two-body correlations. The entropy grows monotonically with $|\lambda|$ reflecting increasing effect of the density-assisted process that entangles particles. The entropy however, is not monotonous in time, it builds up from zero, reaches a peak, dips around $t=\tau/2$ and returns to zero at $t = \tau$. It means that the state comes back to non-correlated state, reflecting the periodic nature of the Bloch dynamics. 

In left panels in Fig.~\ref{fig:entropy_zero_onsite} we show the entropy values as a function of time for $\lambda>0$ and $\lambda<0$ respectively. For small values $|\lambda| \lesssim 0.5$, the entropy is higher in the $\lambda < 0$ regime than in the $\lambda > 0$ regime at the same $|\lambda|$. This general trend holds across most values of $\lambda$, however, a noticeable difference is visible around $\lambda = +3/4$, where the entropy reaches its global maximum $\mathcal{S}_{\max} \approx 2.4$. Although $\mathcal{S}_{\max}$ remains well below the upper bound $\ln(M) \approx 4.6$, which is set by the fully mixed state. Its occurrence in the regime where $J$ and $T$ are of opposite sign is counterintuitive and suggests that the partial cancellation between $J$ and $T$ does not merely suppress the dynamics but instead induces non-trivial two-body correlations. We also note that the tilt $F$ plays no role in the generation of correlations, \textit{i.e.} the entropy $\mathcal{S}$ remains constant as a function of $F$ for all values of $\lambda$, confirming that the external potential only drives the centre-of-mass motion by exhibiting Bloch oscillations without contributing to the two-body correlations. The correlations are generated exclusively by the density-assisted tunnelling, which (for $U=0$) is the only two-body term in the Hamiltonian.
\begin{figure}
    \centering
    \includegraphics[width=\linewidth]{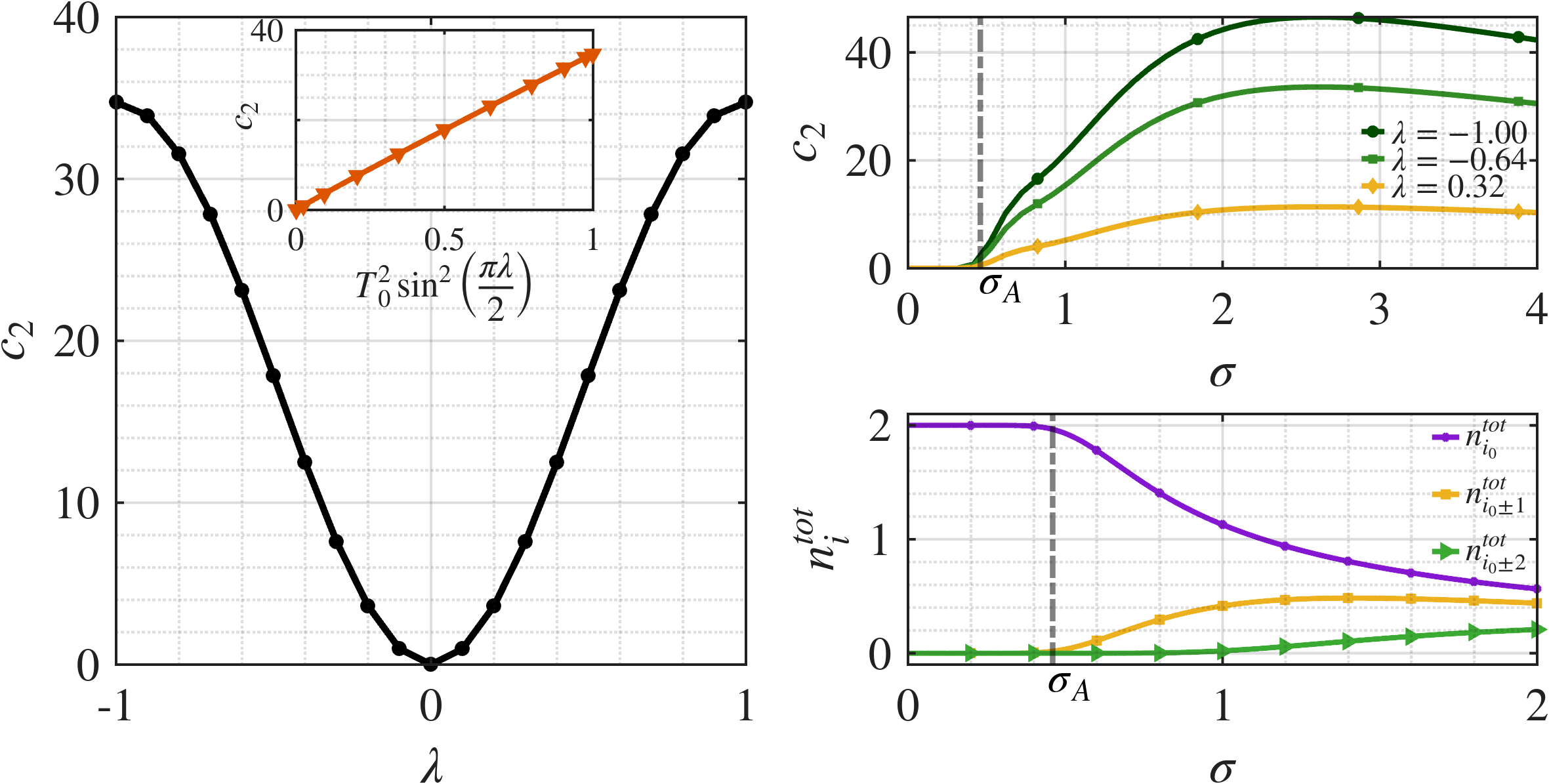}
    \caption{Early-time entropy generation coefficient $c_2$ as a function of $\lambda$ and Gaussian width $\sigma$. Inset shows $c_2$ plotted against $T^2$, collapsing onto a straight line. Bottom-right panel shows the initial-state site occupations $n_{i_0}$, $n_{i_0\pm1}$, and $n_{i_0\pm2}$ at the central, nearest-neighbour, and next-nearest-neighbour sites as a function of $\sigma$.}
    \label{fig:second_derivative}
\end{figure}
%%-------------------------maybe a new section?---------
\section{Entanglement generation at early-times}
\label{sec:earlytime_entanglement}
To establish a direct and quantitative relation between the entropy and the density-assisted tunnelling, we expand the entropy at early times as $\mathcal{S}(t) = c_2t^2+c_4t^4 \dots$. The odd powers vanish by the time reversal symmetry and $ \mathcal{S}(0)=0$ for the chosen initial state. The leading coefficient $c_2$ therefore captures the initial rate of entanglement production and serves as a direct probe for the two-body correlations generated by the density-assisted tunnelling. We extract this coefficient numerically $c_2\approx \frac{\mathcal{S}(t)}{t^2}$ at very early time $t\approx10^{-5}$, which is well withing the perturbative regime $t \ll \tau = \frac{2\pi \hbar}{F}$. We display it in the left panel of Fig.~\ref{fig:second_derivative} as a function of $\lambda$. Immediately it becomes clear that the coefficient $c_2$ follows the same sinusoidal nature encoded in the density-assisted tunnelling vanishing at $\lambda=0$ and increasing monotonously with $|\lambda|>0$. Secondly, when plotted against $T^2$, $c_2$ collapses to a straight line passing through zero indicating $c_2\propto T^2$.

\begin{figure*}
    \centering
    \includegraphics[width=\linewidth]{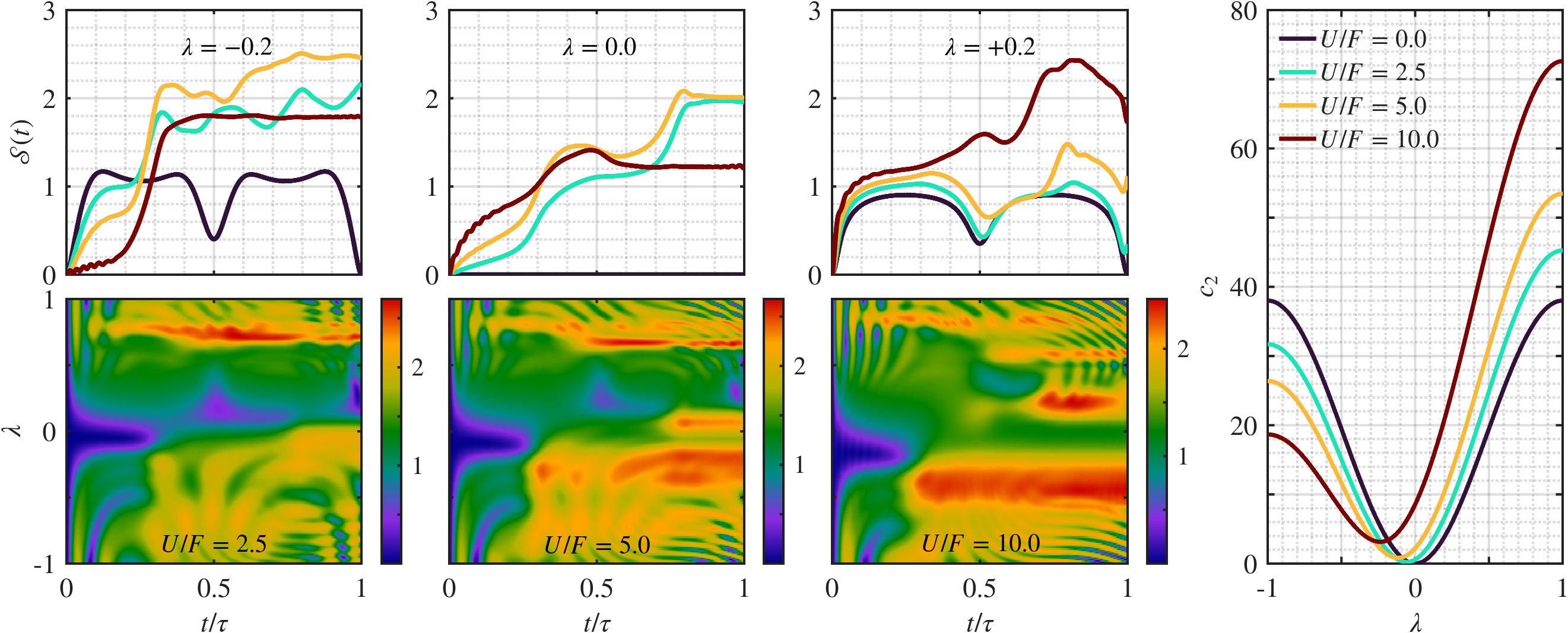}
    \caption{The von Neumann entropy $\mathcal{S}(t)$ for two interacting bosons for different values of the onsite interaction $U$. The top row shows the time evolution of $\mathcal{S}(t)$ over one Bloch period for $\lambda=-0.2$, $0.0$, and $+0.2$, corresponding to the reinforcing, zero density-assisted tunnelling, and competing regimes, respectively. The bottom row shows $\mathcal{S}(t)$ as a function of $\lambda$ and $t/\tau$ for $U/F=2.5$, $5.0$, and $10.0$. The righmost panel shows the early-time entropy generation coefficient $c_2$ at $t\approx10^{-5}$ as a function of $\lambda$ for different interaction strengths.}
    \label{fig:entropy_int}
\end{figure*}
Having established that the $c_2$ is directly proportional to $T^2$, we now probe how $T$ begins to affect the dynamics. To visualise this, we compute how fast the entropy is generated for different initial widths $\sigma$. In the top right panel of Fig.~\ref{fig:second_derivative} we show the entropy coefficient $c_2$ as a function of $\sigma$ for different $\lambda\neq0$ values. A large $c_2$ means that the correlations are built up rapidly at early times. Obviously, for $\lambda = 0$, $c_2\equiv0$, affirming that standard hopping alone generates no inter-particle correlations regardless of the initial state. As soon as $|\lambda| \neq 0$, the correlations begin to build due to density-assisted tunnelling. However, it happens only after $\sigma$ exceeds a threshold value $\sigma_A \approx 0.46$ shown as dashed line. Below this threshold, the initial wavepacket is sufficiently narrow that the probability of finding the two bosons on neighbouring sites is negligible. Consequently no significant inter-particle correlations are generated. Above this threshold, the nearest-neighbour overlap grows, the density-assisted tunnelling becomes effective, and $c_2$ rises rapidly with $\sigma$.
%as an increasing number of two-body states contribute to the dynamics. This increase, however, is not monotonic. The first derivative of $c_2$ with respect to the width $\sigma$ reaches a maximum around $\sigma_B\approx0.55$ (which is shown on the figure as a dotted line).

Phenomenologically, the origin of the threshold $\sigma_A$ can be explained by studying the site occupations $n_{i_0}^{tot}$, $n_{{i_0}\pm1}^{tot}$, and $n_{{i_0}\pm2}^{tot}$ as a function of $\sigma$ (bottom-right panel of Fig.~\ref{fig:second_derivative}). At small $\sigma$, both bosons are localised at the centre, forming a doublon ($n_{i_0}^{tot} \approx 2$ and $n^{tot}_{i_0\pm1}\approx 0$) and the short-time dynamics does not generate entanglement (see Appendix \ref{app:reduced_dynamics}). As $\sigma$ increases beyond the threshold $\sigma_A\approx0.46$ (dashed line), the nearest-neighbour sites begin to populate, and the density-assisted tunnelling generates inter-particle correlations. This onset in $n^{tot}_{{i_0}\pm1}$ coincides directly with the threshold observed in $c_2$, confirming that the threshold $\sigma_A$ is controlled by the spatial extent of the initial wavepacket rather than by the relative strengths of $J$ and $T$. At larger $\sigma$, the initial state spreads over many lattice sites, reducing the local density that the density-assisted tunnelling requires to generate correlations. It is clear that for large enough spreading of the initial state (for example $\sigma\gtrsim1.5$), occupation at second-neighbouring sites ($n^{tot}_{i_0\pm2}$) cannot be neglected. This opens additional possibilities for density-assisted tunnelling and is responsible for the change in slope of $c_2$ which reaches maximum and decreases gradually for even larger $\sigma$. This implies that the initial state has to be wide enough to place the two bosons on neighbouring sites for the entanglement to be generated. Once this condition is met, the correlations grow as $c_2 t^2$, with $c_2 \propto T^2$. 
%%%%%%%%%%%%%%%%%%%%%%%%%%%%%%%%%%%%%%%%%%%%%%%
%--------Interaction section
%%%%%%%%%%%%%%%%%%%%%%%%%%%%%%%%%%%%%%%%%%%%%%%%%%%%%%%%
\section{Effect of onsite interactions on correlations}
\label{sec:onsite_interactions}
So far we have set $U=0$ and studied how the two tunnelling mechanisms $J$ and $T$ drive the centre of mass dynamics and build correlations between the two bosons. We now switch on the onsite interaction and examine how it changes this picture. The case without density-assisted tunnelling, ($\lambda=0$), has been studied in detail in Ref.~\cite{Dariusz:2017}, and we use it here as a reference result. Our main interest is in how a finite $U$ changes the correlations when both tunnelling mechanisms are present. For this purpose, we focus on the von Neumann entropy of the single-particle density matrix $\mathcal{S}(t)$ and choose $\lambda= -0.2, 0, +0.2$ for varying values of the onsite interaction parameter $U$.

We first consider the entropy as a function of time for one Bloch period, shown in the top row of Fig.~\ref{fig:entropy_int}. In the regime where $\lambda=-0.2$, the non-interacting curve returns to zero at the end of the Bloch period recovering the observation made in the previous section. Once the interaction is switched on, the entropy does not return to zero anymore. For smaller values of interaction $U/F=2.5$ and $U/F=5$, the entropy increases with time and does not reach zero in the timescale shown. Increasing the interaction further reduces the entropy which is seen for $U/F=10$, as the probability of doublons forming is suppressed with stronger repulsive interaction. In limit of zero density-assisted tunnelling ($\lambda=0$), the non-interacting particles produce zero entropy as expected. Increasing the onsite interaction in this case generates the particle correlations and again, the entropy does not return to zero at the end of the Bloch cycle. The case of $\lambda=0$ resembles the case of reinforcing regime ($\lambda<0$) with the entropy increasing for smaller values of $U$ and decreasing for larger onsite interaction. The case of the competing regime ($\lambda=0.2$) is qualitatively different from the other two. In this case the entropy for smaller values of $U/F=0,~2.5$ show periodic-like behaviour, but the decrease of entropy for increasing $U$ is no longer observed. Instead the entropy shows a monotonic increase remaining small and periodic-like for weaker interactions $U$, and increasing for large values of $U$. In all three cases, as the interaction is increased, the entropy dynamics develops increasingly rapid oscillations. 

For completeness, we show in the bottom row of Fig.~\ref{fig:entropy_int}, how these changes extend across the full range of $\lambda$ for one Bloch period. Comparing this with the case without onsite interaction in Fig.~\ref{fig:entropy_zero_onsite}, we immediately notice that the entropy is nonzero around $\lambda=0$ when onsite interactions are present. For smaller values of the onsite interaction $U/F=2.5, 5$, the maximum entropy is produced in the competing regime $\lambda>0$. This changes for large interaction strength, where the maximum entropy is instead seen in the reinforcing regime $\lambda<0$.

Lastly, to quantify how the interaction changes the initial generation of correlations, we extract the coefficient $c_2$ at $t\approx10^{-5}$ and plot it as a function of $\lambda$ in the rightmost panel of Fig.~\ref{fig:entropy_int}. In the non-interacting limit, $c_2$ follows the dependence of the density-assisted tunnelling strength and vanishes at $\lambda=0$. Once $U$ is finite, $c_2$ becomes nonzero at $\lambda=0$, showing directly that the onsite interaction provides an additional way for generating correlations. Note also that the balanced nature of the curve is lost, with the entropy production in the competing regime becoming larger than in the reinforcing regime. We see that the effect of the onsite interaction depends strongly on the sign of $\lambda$ and leads to qualitatively different entropy dynamics in the reinforcing and competing regimes.

%%=============lambda 1/2 case--------------
\section{Destructive interference of the tunnellings at \texorpdfstring{$\lambda=1/2$}{lambda=1/2}}
\label{sec:lambda_0.5}

Finally we consider the special case $\lambda=1/2$, where the standard and density-assisted tunnelling interfere destructively. This leads to an exact decoupling of the doublon and singlon subspaces, with direct consequences for the density and entanglement dynamics as discussed below.

Action of the Hamiltonian \eqref{eq:finalHamiltonian} on any doublon state is 
\begin{align}
    \hat H|D_k\rangle &= \sqrt{2}(-J+T)
    \left(|S_{k-1,k}\rangle + |S_{k,k+1}\rangle\right)  \nonumber \\
    &+ U|D_k\rangle + 2(k-i_0)F|D_k\rangle.
    \label{eq:doublon}
\end{align}
Therefore, for $T=J$, ($\lambda=1/2$), there is no coupling between the doublon and the singlon subspaces. This observation has dramatic consequences on the dynamics of the doublon density $n_i^{(2)}(t)$. Since the general state of the system at any time $t$ can be written as a superposition
\begin{align}
    |\Psi(t)\rangle = \sum_k \gamma_k(t) |D_k\rangle + \sum_{k<l} \delta_{kl}(t) |S_{kl}\rangle\;.
\end{align}
One finds
\begin{align}
    i\dot \gamma_k(t) = \sum_l \gamma_l(t) \langle D_k|\hat H|D_l \rangle + \sum_{l<m} \delta_{lm}(t) \langle D_k|\hat H| S_{lm}\rangle\;.
\end{align}
Consequently, for $T=J$, one gets 
\begin{align}
    \dot \gamma_k(t) &= -i \gamma_k(t)\big(U+2F(k-i_0)\big) \nonumber \\
    \gamma_k(t) &= \gamma_0 e^{-i(U +2F(k-i_0))}\;.
\end{align}
It means that $n_k^{(2)}(t) = |\gamma_k(t)|^2 = |\gamma_0|^2$ is time independent.
\begin{figure}[h!]
\centering
\includegraphics[width=\linewidth]{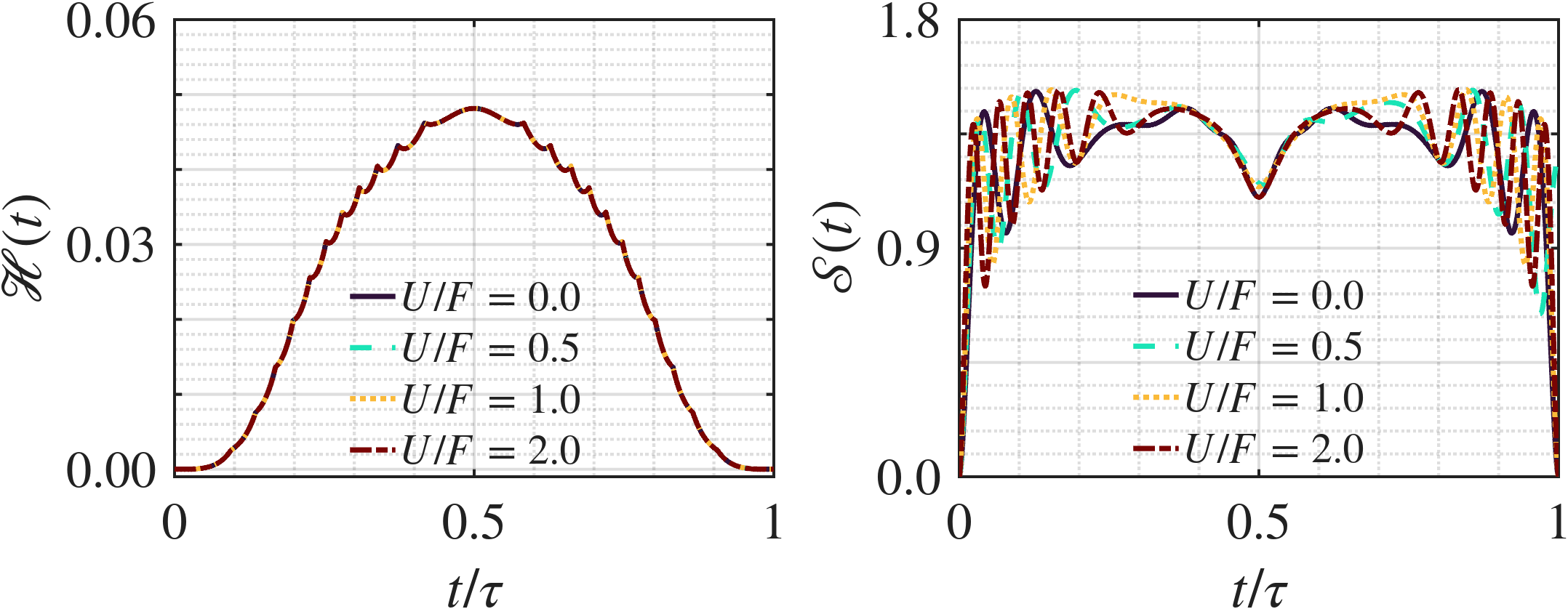}
\caption{Hellinger distance $\mathscr{H}(t)$ and entropy $\mathcal{S}(t)$ over one Bloch period for various interaction strengths $U/F$, at the destructive interference point $\lambda=1/2$. At this value, all curves for the Hellinger distance are indistinguishable, showing that the density dynamics is independent of $U$ at the point $J=T$. The entropy however shows slight deviations with different interaction  values.}
\label{fig:hellinger_entropy_int}
\end{figure}

This separation of the doublon and singlon sectors has a direct consequence on the Hellinger distance and entropy as shown in Fig.~\ref{fig:hellinger_entropy_int}. The doublon amplitudes acquire only an interaction and tilt dependent phase, while their probabilities remain unchanged, $n_i^{(2)}(t)=|\gamma_0|^2$. Consequently, the density dynamics, and hence the Hellinger distance $\mathscr{H}(t)$, is independent of $U$, which explains the complete overlap of the Hellinger curves for different interaction strengths. The entropy $\mathcal{S}(t)$, however, retains a very weak dependence on $U$. 
%%--------------------end of results-------------
\section{Conclusions and outlook}
\label{sec:conclusion}

In this work, we investigated the dynamics of two bosonic quantum walkers in a tilted one-dimensional optical lattice described by an extended Bose--Hubbard model with specific inclusion of density-assisted tunnelling $T$. By varying the relative strength and sign of the two tunnelling processes through the parameter $\lambda$, we showed distinct reinforcing and competing regimes of the dynamical behaviour with qualitatively different density dynamics. In the reinforcing regime, the two tunnelling processes enhance each other and produce larger amplitude Bloch oscillations, while in the competing regime their partial cancellation suppresses the oscillations and can also lead to non-oscillating density components. To quantify the changes in the density profile we used Hellinger distance which highlights the asymmetry between positive and negative $\lambda$.

We also examined the correlations generated by density-assisted tunnelling in the absence of onsite interactions. The correlations in this limit originate from the two-body nature of density-assisted tunnelling. The early-time entropy follows the same behaviour as that of $T$, $\mathcal{S}(t)\simeq c_2t^2$, with the coefficient $c_2\propto T^2$. Moreover, we found that the generation of correlations depends crucially on the spatial extent of the initial wavepacket $\sigma$. For a sufficiently narrow initial Gaussian state, the absence of nearest-neighbour occupation suppresses the early-time entropy generation. Once the wavepacket becomes sufficiently broad to populate neighbouring sites, correlations develop quickly and $c_2$ increases before eventually decreasing for broader wavepackets.

Inclusion of  finite onsite interaction changes this picture by providing an additional mechanism for generating correlations. Its response to increasing $U$ depends strongly on the sign of $\lambda$. We also found a special case of $\lambda=1/2$ corresponding to equal tunnelling amplitudes $J=T$. At this point, the coupling between the doublon and singlon states cancel each other completely. The doublon sector therefore becomes dynamically frozen, while the singlon sector continues to evolve through conventional tunnelling $J$ and exhibits Bloch oscillations. This decoupling is independent of the onsite interaction, lattice tilt, and initial state.

As an outlook, the doublon–singlon decoupling at $J=T$ can be explored further for large particle numbers and with inclusion of pair tunnelling and dipolar interactions \cite{Aramthottil:2023}. Another direction would be to investigate density-assisted tunnelling in non-Hermitian settings by including non-reciprocal hoppings in one and two component systems \cite{Shen:2022, Brighi:2024, Hasan:2026}, where the interference between the two tunnelling channels could lead to new forms of transport of bosons or interaction-induced skin effect. These questions are also relevant experimentally, given demonstration of density-dependent tunnelling and engineered interactions in optical lattices \cite{Jurgensen:2014, Meinert:2016, Gorg:2019}. Density-assisted and correlated hopping have been shown to give rise to quantum many-body scars and slow dynamics \cite{Hudomal:2020}, while interaction-induced tunnelling in dipolar bosons can lead to delocalisation and nonergodic dynamics \cite{Aramthottil:2023}.

\section*{Data availability}
All numerical data presented in this paper are openly available
at \cite{Hasan_zenodo:2026}. 

\section{Acknowledgements}
MSH thanks Abhishek Anand, Tai Tran, Jacek Dobrzyniecki for teaching second quantisation and Salassal for fruitful discussions at the early stages of the project. VM thanks Abhishek Anand for discussions. This research was supported by the National Science Centre (NCN, Poland) within the OPUS project No. 2023/49/B/ST2/03744. For the purpose of Open Access, the authors have applied a CC-BY public copyright licence to any Author Accepted Manuscript version arising from this submission.
%%------------------appendix one column------------
% \clearpage
\appendix
\onecolumngrid
% %%--------------------------
\section{Short-time dynamics of localised doublon}
\label{app:reduced_dynamics}
We consider two bosons initially occupying the central site $i_0$, corresponding to the limit $\sigma\to0$. The initial state is a doublon state.
\begin{equation}
    |\psi(0)\rangle=|D_{i_0}\rangle
    =\frac{(\hat a_{i_0}^{\dagger})^2}{\sqrt{2}}|0\rangle.
\end{equation}
The time-evolved state upto second order in time $t$ is 
\begin{equation}
    |\psi(t)\rangle
    =\left(1-i\hat Ht-\frac{t^2}{2}\hat H^2
    +\mathcal{O}(t^3)\right)|D_{i_0}\rangle\;,
\end{equation}
which gives 
\begin{align}
    |\psi(t)\rangle
    =& \left(1-iUt-\frac{t^2}{2}(U^2+2\gamma^2)\right)|D_{i_0}\rangle
    - \frac{\gamma^2t^2}{2}\left(|D_{i_0-1}\rangle+|D_{i_0+1}\rangle\right)
    \nonumber\\
    +& \left(-i\gamma t + \frac{F\gamma t^2}{2}\right)|S_{i_0-1,i_0}\rangle
    + \left(-i\gamma t - \frac{F\gamma t^2}{2}\right)|S_{i_0,i_0+1}\rangle
    \nonumber\\
    +& \frac{J\gamma t^2}{2}\left(|S_{i_0-2,i_0}\rangle + 2|S_{i_0-1,i_0+1}\rangle + |S_{i_0,i_0+2}\rangle\right)
    + \mathcal O(t^3)\,,
    \label{eq:short_time_state}
\end{align}
where $\gamma = \sqrt{2}(-J+T)$.

The reduced single-particle density matrix is
\begin{equation}
    \hat\rho^{(1)} = \sum_{kl}\rho^{(1)}_{kl}\,|k\rangle\langle l|\,,
    \qquad
    \rho^{(1)}_{kl} = \langle\psi(t)|\hat a^\dagger_{i_0+k}\hat a_{i_0+l}|\psi(t)\rangle\,,
\end{equation}
where $|k\rangle=\hat a^\dagger_{i_0+k}|0\rangle$. By substituting explicit form of \eqref{eq:short_time_state}, we find that the only non-vanishing matrix elements are,
\begin{align}
    \rho^{(1)}_{0,0} &= 2\left(1-\gamma^2t^2\right)\,, \nonumber\\
    \rho^{(1)}_{1,1} &= \rho^{(1)}_{-1,-1} = \rho^{(1)}_{-1,1} = \rho^{(1)}_{1,-1} = \gamma^2t^2\,, \nonumber\\
    \rho^{(1)}_{0,1} &= \sqrt2\left(-i\gamma t + \gamma  \frac{(U-F)}{2} t^2\right) = \Big(\rho^{(1)}_{1,0}\Big)^*\,, \nonumber\\
    \rho^{(1)}_{0,-1} &= \sqrt2\left(-i\gamma t + \gamma  \frac{(U+F)}{2} t^2\right)= \Big(\rho^{(1)}_{-1,0}\Big)^*\,,\nonumber \\    
    \rho^{(1)}_{0,\pm2} &= \rho^{(1)}_{\pm2,0} = \frac{J\gamma}{\sqrt2}\,t^2\,.
    \label{eq:rho_elements}
\end{align}
It can be straightforwardly shown that this density matrix can be expressed with one single--particle orbital 
\begin{equation}
     \rho^{(1)} = 2\; |\phi(t)\rangle\langle\phi(t)|
\end{equation}
where 
\begin{equation}
    |\phi(t)\rangle
    = \sqrt2\left(1-\frac{\gamma^2t^2}{2}\right)|0\rangle
    + \left[i\gamma t+\frac{\gamma(U-F)}{2}t^2\right]|1\rangle
    + \left[i\gamma t+\frac{\gamma(U+F)}{2}t^2\right]|{-1}\rangle
    + \frac{J\gamma t^2}{2}\left(|{-2}\rangle+|2\rangle\right)\;.
    \label{eq:phi}
\end{equation}
Consequently, the von Neumann entropy vanishes in the limit $\sigma\rightarrow 0$ independently on parameters of the Hamiltonian.
\twocolumngrid
%%---------------------references------------------
\bibliography{references}

@article{Zhang:2010,
   title={Exact diagonalization: the Bose–Hubbard model as an example},
   volume={31},
   ISSN={1361-6404},
   url={http://dx.doi.org/10.1088/0143-0807/31/3/016},
   DOI={10.1088/0143-0807/31/3/016},
   number={3},
   journal={European Journal of Physics},
   publisher={IOP Publishing},
   author={Zhang, J M and Dong, R X},
   year={2010},
   month={Apr}, 
   pages={591–602} 
}

@article{Raventos:2017,
   title={Cold bosons in optical lattices: a tutorial for exact diagonalization},
   volume={50},
   ISSN={1361-6455},
   url={http://dx.doi.org/10.1088/1361-6455/aa68b1},
   DOI={10.1088/1361-6455/aa68b1},
   number={11},
   journal={Journal of Physics B: Atomic, Molecular and Optical Physics},
   publisher={IOP Publishing},
   author={Raventós, David and Graß, Tobias and Lewenstein, Maciej and Juliá-Díaz, Bruno},
   year={2017},
   month={May},
   pages={113001} 
}

@article{Fisher:1989,
  title = {Boson localization and the superfluid-insulator transition},
  author = {Fisher, Matthew P. A. and Weichman, Peter B. and Grinstein, G. and Fisher, Daniel S.},
  journal = {Phys. Rev. B},
  volume = {40},
  issue = {1},
  pages = {546--570},
  numpages = {0},
  year = {1989},
  month = {Jul},
  publisher = {American Physical Society},
  doi = {10.1103/PhysRevB.40.546},
  url = {https://link.aps.org/doi/10.1103/PhysRevB.40.546}
}

@article{Dariusz:2017,
  title = {Two bosonic quantum walkers in one-dimensional optical lattices},
  author = {Wiater, Dariusz and Sowi\ifmmode \acute{n}\else \'{n}\fi{}ski, Tomasz and Zakrzewski, Jakub},
  journal = {Phys. Rev. A},
  volume = {96},
  issue = {4},
  pages = {043629},
  numpages = {11},
  year = {2017},
  month = {Oct},
  publisher = {American Physical Society},
  doi = {10.1103/PhysRevA.96.043629},
  url = {https://link.aps.org/doi/10.1103/PhysRevA.96.043629}
}

@article{Sarkar:2020,
  title = {Correlations in few two-component quantum walkers on a tilted lattice},
  author = {Sarkar, Saubhik and Sowi\ifmmode \acute{n}\else \'{n}\fi{}ski, Tomasz},
  journal = {Phys. Rev. A},
  volume = {102},
  issue = {4},
  pages = {043326},
  numpages = {7},
  year = {2020},
  month = {Oct},
  publisher = {American Physical Society},
  doi = {10.1103/PhysRevA.102.043326},
  url = {https://link.aps.org/doi/10.1103/PhysRevA.102.043326}
}

@article{Sowinski:2012,
  title = {Dipolar Molecules in Optical Lattices},
  author = {Sowi\ifmmode \acute{n}\else \'{n}\fi{}ski, Tomasz and Dutta, Omjyoti and Hauke, Philipp and Tagliacozzo, Luca and Lewenstein, Maciej},
  journal = {Phys. Rev. Lett.},
  volume = {108},
  issue = {11},
  pages = {115301},
  numpages = {5},
  year = {2012},
  month = {Mar},
  publisher = {American Physical Society},
  doi = {10.1103/PhysRevLett.108.115301},
  url = {https://link.aps.org/doi/10.1103/PhysRevLett.108.115301}
}

@article{Mazzarella:2006,
  title = {Extended Bose Hubbard model of interacting bosonic atoms in optical lattices: From superfluidity to density waves},
  author = {Mazzarella, G. and Giampaolo, S. M. and Illuminati, F.},
  journal = {Phys. Rev. A},
  volume = {73},
  issue = {1},
  pages = {013625},
  numpages = {15},
  year = {2006},
  month = {Jan},
  publisher = {American Physical Society},
  doi = {10.1103/PhysRevA.73.013625},
  url = {https://link.aps.org/doi/10.1103/PhysRevA.73.013625}
}

@article{Jurgensen:2014,
  title = {Observation of Density-Induced Tunneling},
  author = {J\"urgensen, Ole and Meinert, Florian and Mark, Manfred J. and N\"agerl, Hanns-Christoph and L\"uhmann, Dirk-S\"oren},
  journal = {Phys. Rev. Lett.},
  volume = {113},
  issue = {19},
  pages = {193003},
  numpages = {5},
  year = {2014},
  month = {Nov},
  publisher = {American Physical Society},
  doi = {10.1103/PhysRevLett.113.193003},
  url = {https://link.aps.org/doi/10.1103/PhysRevLett.113.193003}
}

@article{Meinert:2016,
   title={Floquet Engineering of Correlated Tunneling in the Bose-Hubbard Model with Ultracold Atoms},
   volume={116},
   ISSN={1079-7114},
   url={http://dx.doi.org/10.1103/PhysRevLett.116.205301},
   DOI={10.1103/physrevlett.116.205301},
   number={20},
   journal={Physical Review Letters},
   publisher={American Physical Society (APS)},
   author={Meinert, F. and Mark, M.J. and Lauber, K. and Daley, A.J. and Nägerl, H.-C.},
   year={2016},
   month={May}
}

@article{Jurgensen:2012,
  title = {Density-induced processes in quantum gas mixtures in optical lattices},
  author = {J\"urgensen, Ole and Sengstock, Klaus and L\"uhmann, Dirk-S\"oren},
  journal = {Phys. Rev. A},
  volume = {86},
  issue = {4},
  pages = {043623},
  numpages = {10},
  year = {2012},
  month = {Oct},
  publisher = {American Physical Society},
  doi = {10.1103/PhysRevA.86.043623},
  url = {https://link.aps.org/doi/10.1103/PhysRevA.86.043623}
}

@article{Heinze:2011,
  title = {Multiband Spectroscopy of Ultracold Fermions: Observation of Reduced Tunneling in Attractive Bose-Fermi Mixtures},
  author = {Heinze, J. and G\"otze, S. and Krauser, J. S. and Hundt, B. and Fl\"aschner, N. and L\"uhmann, D.-S. and Becker, C. and Sengstock, K.},
  journal = {Phys. Rev. Lett.},
  volume = {107},
  issue = {13},
  pages = {135303},
  numpages = {5},
  year = {2011},
  month = {Sep},
  publisher = {American Physical Society},
  doi = {10.1103/PhysRevLett.107.135303},
  url = {https://link.aps.org/doi/10.1103/PhysRevLett.107.135303}
}

@article{Ospelkaus:2006,
  title = {Localization of Bosonic Atoms by Fermionic Impurities in a Three-Dimensional Optical Lattice},
  author = {Ospelkaus, S. and Ospelkaus, C. and Wille, O. and Succo, M. and Ernst, P. and Sengstock, K. and Bongs, K.},
  journal = {Phys. Rev. Lett.},
  volume = {96},
  issue = {18},
  pages = {180403},
  numpages = {4},
  year = {2006},
  month = {May},
  publisher = {American Physical Society},
  doi = {10.1103/PhysRevLett.96.180403},
  url = {https://link.aps.org/doi/10.1103/PhysRevLett.96.180403}
}

@article{Gunter:2006,
  title = {Bose-Fermi Mixtures in a Three-Dimensional Optical Lattice},
  author = {G\"unter, Kenneth and St\"oferle, Thilo and Moritz, Henning and K\"ohl, Michael and Esslinger, Tilman},
  journal = {Phys. Rev. Lett.},
  volume = {96},
  issue = {18},
  pages = {180402},
  numpages = {4},
  year = {2006},
  month = {May},
  publisher = {American Physical Society},
  doi = {10.1103/PhysRevLett.96.180402},
  url = {https://link.aps.org/doi/10.1103/PhysRevLett.96.180402}
}

@article{Dias:2007,
  title = {Frequency doubling of Bloch oscillations for interacting electrons in a static electric field},
  author = {Dias, W. S. and Nascimento, E. M. and Lyra, M. L. and de Moura, F. A. B. F.},
  journal = {Phys. Rev. B},
  volume = {76},
  issue = {15},
  pages = {155124},
  numpages = {6},
  year = {2007},
  month = {Oct},
  publisher = {American Physical Society},
  doi = {10.1103/PhysRevB.76.155124},
  url = {https://link.aps.org/doi/10.1103/PhysRevB.76.155124}
}

@article{Hudomal:2020,
   title={Quantum scars of bosons with correlated hopping},
   volume={3},
   ISSN={2399-3650},
   url={http://dx.doi.org/10.1038/s42005-020-0364-9},
   DOI={10.1038/s42005-020-0364-9},
   number={1},
   journal={Communications Physics},
   publisher={Springer Science and Business Media LLC},
   author={Hudomal, Ana and Vasic, Ivana and Regnault, Nicolas and Papic, Zlatko},
   year={2020},
   month={June}
}

@article{Khomeriki:2010,
  title = {Interaction-induced fractional Bloch and tunneling oscillations},
  author = {Khomeriki, Ramaz and Krimer, Dmitry O. and Haque, Masudul and Flach, Sergej},
  journal = {Phys. Rev. A},
  volume = {81},
  issue = {6},
  pages = {065601},
  numpages = {4},
  year = {2010},
  month = {Jun},
  publisher = {American Physical Society},
  doi = {10.1103/PhysRevA.81.065601},
  url = {https://link.aps.org/doi/10.1103/PhysRevA.81.065601}
}

@article{Rabec:2025,
   title={Bloch oscillations of a soliton in a one-dimensional quantum fluid},
   volume={21},
   ISSN={1745-2481},
   url={http://dx.doi.org/10.1038/s41567-025-02970-1},
   DOI={10.1038/s41567-025-02970-1},
   number={10},
   journal={Nature Physics},
   publisher={Springer Science and Business Media LLC},
   author={Rabec, F. and Chauveau, G. and Brochier, G. and Nascimbene, S. and Dalibard, J. and Beugnon, J.},
   year={2025},
   month={aug},
   pages={1541–1547} 
}

@article{Hasan:2022,
doi = {10.1088/1361-6455/ac6ea3},
url = {https://doi.org/10.1088/1361-6455/ac6ea3},
year = {2022},
month = {may},
publisher = {IOP Publishing},
volume = {55},
number = {13},
pages = {135302},
author = {Hasan, Muhammad S and Polo, J and Pelayo, J C and Busch, Th},
title = {Bloch oscillations in supersolids},
journal = {Journal of Physics B: Atomic, Molecular and Optical Physics},
}

@article{Morsch:2016,
doi = {10.1088/1367-2630/18/11/111005},
url = {https://doi.org/10.1088/1367-2630/18/11/111005},
year = {2016},
month = {nov},
publisher = {IOP Publishing},
volume = {18},
number = {11},
pages = {111005},
author = {Morsch, Oliver},
title = {Taking a peek at Bloch oscillations},
journal = {New Journal of Physics},
}

@article{Jaksch:1998,
  title = {Cold Bosonic Atoms in Optical Lattices},
  author = {Jaksch, D. and Bruder, C. and Cirac, J. I. and Gardiner, C. W. and Zoller, P.},
  journal = {Phys. Rev. Lett.},
  volume = {81},
  issue = {15},
  pages = {3108--3111},
  numpages = {0},
  year = {1998},
  month = {Oct},
  publisher = {American Physical Society},
  doi = {10.1103/PhysRevLett.81.3108},
  url = {https://link.aps.org/doi/10.1103/PhysRevLett.81.3108}
}

@article{Chanda:2025,
doi = {10.1088/1361-6633/adc3a7},
url = {https://doi.org/10.1088/1361-6633/adc3a7},
year = {2025},
month = {apr},
publisher = {IOP Publishing},
volume = {88},
number = {4},
pages = {044501},
author = {Chanda, Titas and Barbiero, Luca and Lewenstein, Maciej and Mark, Manfred J and Zakrzewski, Jakub},
title = {Recent progress on quantum simulations of non-standard Bose–Hubbard models},
journal = {Reports on Progress in Physics}
}

@article{Gross:2017,
author = {Christian Gross  and Immanuel Bloch },
title = {Quantum simulations with ultracold atoms in optical lattices},
journal = {Science},
volume = {357},
number = {6355},
pages = {995-1001},
year = {2017},
doi = {10.1126/science.aal3837},
URL = {https://www.science.org/doi/abs/10.1126/science.aal3837}
}

@article{Bloch:2008,
  title = {Many-body physics with ultracold gases},
  author = {Bloch, Immanuel and Dalibard, Jean and Zwerger, Wilhelm},
  journal = {Rev. Mod. Phys.},
  volume = {80},
  issue = {3},
  pages = {885--964},
  numpages = {0},
  year = {2008},
  month = {Jul},
  publisher = {American Physical Society},
  doi = {10.1103/RevModPhys.80.885},
  url = {https://link.aps.org/doi/10.1103/RevModPhys.80.885}
}

@article{Schafer:2020,
   title={Tools for quantum simulation with ultracold atoms in optical lattices},
   volume={2},
   ISSN={2522-5820},
   url={http://dx.doi.org/10.1038/s42254-020-0195-3},
   DOI={10.1038/s42254-020-0195-3},
   number={8},
   journal={Nature Reviews Physics},
   publisher={Springer Science and Business Media LLC},
   author={Schäfer, Florian and Fukuhara, Takeshi and Sugawa, Seiji and Takasu, Yosuke and Takahashi, Yoshiro},
   year={2020},
   month={jul},
   pages={411–425} 
}

@article{Takahashi:2022,
  title={Quantum simulation of quantum many-body systems with ultracold two-electron atoms in an optical lattice},
  author={Yoshiro TAKAHASHI},
  journal={Proceedings of the Japan Academy, Series B},
  volume={98},
  number={4},
  pages={141-160},
  year={2022},
  doi={10.2183/pjab.98.010}
}

@article{Greiner:2002,
   title={Quantum phase transition from a superfluid to a Mott insulator in a gas of ultracold atoms},
   volume={415},
   ISSN={1476-4687},
   url={http://dx.doi.org/10.1038/415039a},
   DOI={10.1038/415039a},
   number={6867},
   journal={Nature},
   publisher={Springer Science and Business Media LLC},
   author={Greiner, Markus and Mandel, Olaf and Esslinger, Tilman and Hänsch, Theodor W. and Bloch, Immanuel},
   year={2002},
   month={jan},
   pages={39–44} 
}

@article{Jordens:2008,
   title={A Mott insulator of fermionic atoms in an optical lattice},
   volume={455},
   ISSN={1476-4687},
   url={http://dx.doi.org/10.1038/nature07244},
   DOI={10.1038/nature07244},
   number={7210},
   journal={Nature},
   publisher={Springer Science and Business Media LLC},
   author={Jördens, Robert and Strohmaier, Niels and Günter, Kenneth and Moritz, Henning and Esslinger, Tilman},
   year={2008},
   month={sep},
   pages={204–207} 
}

@article{Lewenstein:2007,
   title={Ultracold atomic gases in optical lattices: mimicking condensed matter physics and beyond},
   volume={56},
   ISSN={1460-6976},
   url={http://dx.doi.org/10.1080/00018730701223200},
   DOI={10.1080/00018730701223200},
   number={2},
   journal={Advances in Physics},
   publisher={Informa UK Limited},
   author={Lewenstein, Maciej and Sanpera, Anna and Ahufinger, Veronica and Damski, Bogdan and Sen(De), Aditi and Sen, Ujjwal},
   year={2007},
   month={mar},
   pages={243–379} 
}

@ARTICLE{Bloch:1929,
       author = {{Bloch}, Felix},
        title = "{{\"U}ber die Quantenmechanik der Elektronen in Kristallgittern}",
      journal = {Zeitschrift fur Physik},
         year = 1929,
        month = {jul},
       volume = {52},
       number = {7-8},
        pages = {555-600},
          doi = {10.1007/BF01339455} 
}

@article{Morsch:2006,
  title = {Dynamics of Bose-Einstein condensates in optical lattices},
  author = {Morsch, Oliver and Oberthaler, Markus},
  journal = {Rev. Mod. Phys.},
  volume = {78},
  issue = {1},
  pages = {179--215},
  numpages = {0},
  year = {2006},
  month = {Feb},
  publisher = {American Physical Society},
  doi = {10.1103/RevModPhys.78.179},
  url = {https://link.aps.org/doi/10.1103/RevModPhys.78.179}
}

@article{Kolovsky:2004,
author = {KOLOVSKY, ANDREY R. and KORSCH, HANS J{\"U}RGEN},
title = {BLOCH OSCILLATIONS OF COLD ATOMS IN OPTICAL LATTICES},
journal = {International Journal of Modern Physics B},
volume = {18},
number = {09},
pages = {1235-1260},
year = {2004},
doi = {10.1142/S0217979204024483}
}

@article{Dahan:1996,
  title = {Bloch Oscillations of Atoms in an Optical Potential},
  author = {Ben Dahan, Maxime and Peik, Ekkehard and Reichel, Jakob and Castin, Yvan and Salomon, Christophe},
  journal = {Phys. Rev. Lett.},
  volume = {76},
  issue = {24},
  pages = {4508--4511},
  numpages = {0},
  year = {1996},
  month = {Jun},
  publisher = {American Physical Society},
  doi = {10.1103/PhysRevLett.76.4508},
  url = {https://link.aps.org/doi/10.1103/PhysRevLett.76.4508}
}

@article{Chung:1989,
title = {Measures of distance between probability distributions},
journal = {Journal of Mathematical Analysis and Applications},
volume = {138},
number = {1},
pages = {280-292},
year = {1989},
issn = {0022-247X},
doi = {https://doi.org/10.1016/0022-247X(89)90335-1},
url = {https://www.sciencedirect.com/science/article/pii/0022247X89903351},
author = {J.K Chung and P.L Kannappan and C.T Ng and P.K Sahoo}
}

@Article{Vegelius:1986,
author={Vegelius, Jan
and Janson, Svante
and Johansson, Folke},
title={Measures of similarity between distributions},
journal={Quality and Quantity},
year={1986},
month={Dec},
day={01},
volume={20},
number={4},
pages={437-441},
issn={1573-7845},
doi={10.1007/BF00123091},
url={https://doi.org/10.1007/BF00123091}
}

@article{Luigi:2008,
  title = {Entanglement in many-body systems},
  author = {Amico, Luigi and Fazio, Rosario and Osterloh, Andreas and Vedral, Vlatko},
  journal = {Rev. Mod. Phys.},
  volume = {80},
  issue = {2},
  pages = {517--576},
  numpages = {0},
  year = {2008},
  month = {May},
  publisher = {American Physical Society},
  doi = {10.1103/RevModPhys.80.517},
  url = {https://link.aps.org/doi/10.1103/RevModPhys.80.517}
}

@article{Horodecki:2009,
  title = {Quantum entanglement},
  author = {Horodecki, Ryszard and Horodecki, Pawe\l{} and Horodecki, Micha\l{} and Horodecki, Karol},
  journal = {Rev. Mod. Phys.},
  volume = {81},
  issue = {2},
  pages = {865--942},
  numpages = {0},
  year = {2009},
  month = {Jun},
  publisher = {American Physical Society},
  doi = {10.1103/RevModPhys.81.865},
  url = {https://link.aps.org/doi/10.1103/RevModPhys.81.865}
}

@article{Dellanna:2022,
  title = {Entanglement properties and ground-state statistics of free bosons},
  author = {Dell'Anna, Luca},
  journal = {Phys. Rev. A},
  volume = {105},
  issue = {3},
  pages = {032412},
  numpages = {9},
  year = {2022},
  month = {Mar},
  publisher = {American Physical Society},
  doi = {10.1103/PhysRevA.105.032412},
  url = {https://link.aps.org/doi/10.1103/PhysRevA.105.032412}
}

@article{Dur:2002,
  title = {Quantum walks in optical lattices},
  author = {D{\"u}r, W. and Raussendorf, R. and Kendon, V. M. and Briegel, H.-J.},
  journal = {Phys. Rev. A},
  volume = {66},
  issue = {5},
  pages = {052319},
  numpages = {8},
  year = {2002},
  month = {Nov},
  publisher = {American Physical Society},
  doi = {10.1103/PhysRevA.66.052319},
  url = {https://link.aps.org/doi/10.1103/PhysRevA.66.052319}
}

@article{Preiss:2015,
author = {Philipp M. Preiss  and Ruichao Ma  and M. Eric Tai  and Alexander Lukin  and Matthew Rispoli  and Philip Zupancic  and Yoav Lahini  and Rajibul Islam  and Markus Greiner },
title = {Strongly correlated quantum walks in optical lattices},
journal = {Science},
volume = {347},
number = {6227},
pages = {1229-1233},
year = {2015},
doi = {10.1126/science.1260364}
}

@article{Witthaut:2010,
   title={Quantum walks and quantum simulations with Bloch-oscillating spinor atoms},
   volume={82},
   ISSN={1094-1622},
   url={http://dx.doi.org/10.1103/PhysRevA.82.033602},
   DOI={10.1103/physreva.82.033602},
   number={3},
   journal={Physical Review A},
   publisher={American Physical Society (APS)},
   author={Witthaut, D.},
   year={2010},
   month={Sept} 
}

@article{Wang:2014,
  title = {Quantum walks of two interacting anyons in one-dimensional optical lattices},
  author = {Wang, Limin and Wang, Li and Zhang, Yunbo},
  journal = {Phys. Rev. A},
  volume = {90},
  issue = {6},
  pages = {063618},
  numpages = {6},
  year = {2014},
  month = {Dec},
  publisher = {American Physical Society},
  doi = {10.1103/PhysRevA.90.063618},
  url = {https://link.aps.org/doi/10.1103/PhysRevA.90.063618}
}

@article{Karski:2009,
author = {Michal Karski  and Leonid Förster  and Jai-Min Choi  and Andreas Steffen  and Wolfgang Alt  and Dieter Meschede  and Artur Widera },
title = {Quantum Walk in Position Space with Single Optically Trapped Atoms},
journal = {Science},
volume = {325},
number = {5937},
pages = {174-177},
year = {2009},
doi = {10.1126/science.1174436}
}

@article{Eckert:2005,
  title = {One- and two-dimensional quantum walks in arrays of optical traps},
  author = {Eckert, K. and Mompart, J. and Birkl, G. and Lewenstein, M.},
  journal = {Phys. Rev. A},
  volume = {72},
  issue = {1},
  pages = {012327},
  numpages = {9},
  year = {2005},
  month = {Jul},
  publisher = {American Physical Society},
  doi = {10.1103/PhysRevA.72.012327},
  url = {https://link.aps.org/doi/10.1103/PhysRevA.72.012327}
}

@article{Chandrashekar:2008,
   title={Quantum phase transition using quantum walks in an optical lattice},
   volume={78},
   ISSN={1094-1622},
   url={http://dx.doi.org/10.1103/PhysRevA.78.022314},
   DOI={10.1103/physreva.78.022314},
   number={2},
   journal={Physical Review A},
   publisher={American Physical Society (APS)},
   author={Chandrashekar, C. M. and Laflamme, Raymond},
   year={2008},
   month={Aug}
   }

@article{Jaksch:2005,
   title={The cold atom Hubbard toolbox},
   volume={315},
   ISSN={0003-4916},
   url={http://dx.doi.org/10.1016/j.aop.2004.09.010},
   DOI={10.1016/j.aop.2004.09.010},
   number={1},
   journal={Annals of Physics},
   publisher={Elsevier BV},
   author={Jaksch, D. and Zoller, P.},
   year={2005},
   month={Jan},
   pages={52–79} 
   
}

@article{Kuhner:1998,
  title = {Phases of the one-dimensional Bose-Hubbard model},
  author = {K{\"u}hner, T. D. and Monien, H.},
  journal = {Phys. Rev. B},
  volume = {58},
  issue = {22},
  pages = {R14741(R)--R14744(R)},
  numpages = {0},
  year = {1998},
  month = {Dec},
  publisher = {American Physical Society},
  doi = {10.1103/PhysRevB.58.R14741},
  url = {https://link.aps.org/doi/10.1103/PhysRevB.58.R14741}
}

@article{Isart:2007,
   title={Transport and entanglement generation in the Bose–Hubbard model},
   volume={40},
   ISSN={1751-8121},
   url={http://dx.doi.org/10.1088/1751-8113/40/28/S11},
   DOI={10.1088/1751-8113/40/28/s11},
   number={28},
   journal={Journal of Physics A: Mathematical and Theoretical},
   publisher={IOP Publishing},
   author={Romero-Isart, O and Eckert, K and Rodó, C and Sanpera, A},
   year={2007},
   month={June},
   pages={8019–8031} 
   }

@article{Maik:2013,
doi = {10.1088/1367-2630/15/11/113041},
url = {https://doi.org/10.1088/1367-2630/15/11/113041},
year = {2013},
month = {nov},
publisher = {IOP Publishing},
volume = {15},
number = {11},
pages = {113041},
author = {Maik, Michał and Hauke, Philipp and Dutta, Omjyoti and Lewenstein, Maciej and Zakrzewski, Jakub},
title = {Density-dependent tunneling in the extended Bose–Hubbard model},
journal = {New Journal of Physics}
}

@article{Shen:2022,
   title={Non-Hermitian skin clusters from strong interactions},
   volume={5},
   ISSN={2399-3650},
   url={http://dx.doi.org/10.1038/s42005-022-01015-w},
   DOI={10.1038/s42005-022-01015-w},
   number={1},
   journal={Communications Physics},
   publisher={Springer Science and Business Media LLC},
   author={Shen, Ruizhe and Lee, Ching Hua},
   year={2022},
   month={Sept}
}

@article{Brighi:2024,
   title={Nonreciprocal dynamics and the non-Hermitian skin effect of repulsively bound pairs},
   volume={110},
   ISSN={2469-9934},
   url={http://dx.doi.org/10.1103/PhysRevA.110.L020201},
   DOI={10.1103/physreva.110.l020201},
   number={2},
   journal={Physical Review A},
   publisher={American Physical Society (APS)},
   author={Brighi, Pietro and Nunnenkamp, Andreas},
   year={2024},
   month={Aug} 
}

@article{Aramthottil:2023,
   title={Role of interaction-induced tunneling in the dynamics of polar lattice bosons},
   volume={107},
   ISSN={2469-9969},
   url={http://dx.doi.org/10.1103/PhysRevB.107.104305},
   DOI={10.1103/physrevb.107.104305},
   number={10},
   journal={Physical Review B},
   publisher={American Physical Society (APS)},
   author={Aramthottil, Adith Sai and Lacki, Mateusz and Santos, Luis and Zakrzewski, Jakub},
   year={2023},
   month={Mar}
   }

@article{Dutta:2011,
doi = {10.1088/1367-2630/13/2/023019},
url = {https://doi.org/10.1088/1367-2630/13/2/023019},
year = {2011},
month = {feb},
publisher = {},
volume = {13},
number = {2},
pages = {023019},
author = {Dutta, O and Eckardt, A and Hauke, P and Malomed, B and Lewenstein, M},
title = {Bose–Hubbard model with occupation-dependent parameters},
journal = {New Journal of Physics}
}

@article{Luhmann:2012,
doi = {10.1088/1367-2630/14/3/033021},
url = {https://doi.org/10.1088/1367-2630/14/3/033021},
year = {2012},
month = {mar},
publisher = {IOP Publishing},
volume = {14},
number = {3},
pages = {033021},
author = {Lühmann, Dirk-Sören and Jürgensen, Ole and Sengstock, Klaus},
title = {Multi-orbital and density-induced tunneling of bosons in optical lattices},
journal = {New Journal of Physics}
}

@article{Lacki:2013,
doi = {10.1088/1367-2630/15/1/013062},
url = {https://doi.org/10.1088/1367-2630/15/1/013062},
year = {2013},
month = {jan},
publisher = {IOP Publishing},
volume = {15},
number = {1},
pages = {013062},
author = {Lacki, Mateusz and Delande, Dominique and Zakrzewski, Jakub},
title = {Dynamics of cold bosons in optical lattices: effects of higher Bloch bands},
journal = {New Journal of Physics}
}

@article{Bissbort:2012,
   title={Effective multibody-induced tunneling and interactions in the Bose-Hubbard model of the lowest dressed band of an optical lattice},
   volume={86},
   ISSN={1094-1622},
   url={http://dx.doi.org/10.1103/PhysRevA.86.023617},
   DOI={10.1103/physreva.86.023617},
   number={2},
   journal={Physical Review A},
   publisher={American Physical Society (APS)},
   author={Bissbort, Ulf and Deuretzbacher, Frank and Hofstetter, Walter},
   year={2012},
   month={Aug} }

@article{Wang:2019,
title = {Benchmark study of the SF-MI phase transition of the Bose–Hubbard model with density-induced tunneling},
journal = {Physics Letters A},
volume = {383},
number = {7},
pages = {680-683},
year = {2019},
issn = {0375-9601},
doi = {https://doi.org/10.1016/j.physleta.2019.01.012},
url = {https://www.sciencedirect.com/science/article/pii/S0375960119300179},
author = {Tao Wang and Tao Ying}
}

@article{Andraca:2012,
   title={Quantum walks: a comprehensive review},
   volume={11},
   ISSN={1573-1332},
   url={http://dx.doi.org/10.1007/s11128-012-0432-5},
   DOI={10.1007/s11128-012-0432-5},
   number={5},
   journal={Quantum Information Processing},
   publisher={Springer Science and Business Media LLC},
   author={Venegas-Andraca, Salvador Elías},
   year={2012},
   month={July}, pages={1015–1106} 
}

@article{Kempe:2003,
   title={Quantum random walks: An introductory overview},
   volume={44},
   ISSN={1366-5812},
   url={http://dx.doi.org/10.1080/00107151031000110776},
   DOI={10.1080/00107151031000110776},
   number={4},
   journal={Contemporary Physics},
   publisher={Informa UK Limited},
   author={Kempe, J},
   year={2003},
   month={July}, pages={307–327} 
}

@article{Gorg:2019,
   title={Realization of density-dependent Peierls phases to engineer quantized gauge fields coupled to ultracold matter},
   volume={15},
   ISSN={1745-2481},
   url={http://dx.doi.org/10.1038/s41567-019-0615-4},
   DOI={10.1038/s41567-019-0615-4},
   number={11},
   journal={Nature Physics},
   publisher={Springer Science and Business Media LLC},
   author={Görg, Frederik and Sandholzer, Kilian and Minguzzi, Joaquín and Desbuquois, Rémi and Messer, Michael and Esslinger, Tilman},
   year={2019},
   month={Aug}, pages={1161–1167} 
}

@misc{Hasan:2026,
      title={Mobile-impurity dynamics in a non-Hermitian Stark-localised bath}, 
      author={Muhammad S. Hasan and Hsiang-Hua Jen and Jhih-Shih You},
      year={2026},
      eprint={2609.33275},
      archivePrefix={arXiv},
      primaryClass={cond-mat.quant-gas},
      url={https://arxiv.org/abs/2609.33275}, 
}

@article{Kraus:2022,
  title = {Quantum phases of dipolar bosons in one-dimensional optical lattices},
  author = {Kraus, Rebecca and Chanda, Titas and Zakrzewski, Jakub and Morigi, Giovanna},
  journal = {Phys. Rev. B},
  volume = {106},
  issue = {3},
  pages = {035144},
  numpages = {10},
  year = {2022},
  month = {Jul},
  publisher = {American Physical Society},
  doi = {10.1103/PhysRevB.106.035144},
  url = {https://link.aps.org/doi/10.1103/PhysRevB.106.035144}
}

@article{Kolovsky:2003,
  title = {Bloch oscillations of cold atoms in two-dimensional optical lattices},
  author = {Kolovsky, A. R. and Korsch, H. J.},
  journal = {Phys. Rev. A},
  volume = {67},
  issue = {6},
  pages = {063601},
  numpages = {9},
  year = {2003},
  month = {Jun},
  publisher = {American Physical Society},
  doi = {10.1103/PhysRevA.67.063601},
  url = {https://link.aps.org/doi/10.1103/PhysRevA.67.063601}
}

@article{Jaczewski:2024,
   title={Lissajous dynamics of a quantum particle in a tilted two-dimensional discrete lattice},
   volume={139},
   ISSN={2190-5444},
   DOI={10.1140/epjp/s13360-024-05787-8},
   number={11},
   journal={The European Physical Journal Plus},
   publisher={Springer Science and Business Media LLC},
   author={Jaczewski, Grzegorz and Sowiński, Tomasz},
   year={2024},
   month={Nov} 
}

@dataset{Hasan_zenodo:2026,
  author       = {Muhammad Sirajul Hasan and
                  Mittal, Vikash and
                  Sowiński, Tomasz},
  title        = {Supplemental Material for ''{D}ynamics of two bosons
                   with density-assisted tunnelling in a one-
                   dimensional tilted lattice "
                  },
  month        = {oct},
  year         = 2026,
  publisher    = {Zenodo},
  doi          = {10.5281/zenodo.23447679},
  url          = {https://doi.org/10.5281/zenodo.23447679},
}

\end{document}